\documentclass[final,3p,times,twocolumn]{elsarticle}
\usepackage{gensymb}
\usepackage{xcolor}
\usepackage{comment}
\usepackage{placeins}  

\begin{document}
\begin{frontmatter}

\title{Development and Implementation of Advanced Beam Diagnostic \\and Abort Systems in SuperKEKB}

\author[uhm,kmi]{K. Yoshihara\corref{cor1}}
\ead{kyoshiha@hawaii.edu}
\cortext[cor1]{Corresponding author.}
\author[kek,sokendai]{T.~Abe}
\author[nagoya]{M.~Aversano}
\author[cincinnati]{A.~Gale}
\author[kek,sokendai]{H.~Ikeda}
\author[kek,sokendai]{H.~Kaji}
\author[tmu]{H. Kakuno}
\author[tmu]{K. Kitamura}
\author[kek,sokendai]{T.~Koga}
\author[kek,nagoya,kmi]{T.~Iijima}
\author[utokyo]{S.~Kato}
\author[nara]{A.~Kusudo}
\author[sokendai]{Y.~Liu}
\author[nagoya]{A.~Maeda}
\author[iowa]{S.~Mitra}
\author[kek,sokendai]{G.~Mitsuka}
\author[kek,sokendai]{K.~Miyabayashi}
\author[kek,sokendai]{I.~Nakamura}
\author[kek,sokendai]{H.~Nakayama}
\author[kek,sokendai]{Y.~Nakazawa}
\author[utokyo]{R.~Nomaru}
\author[nara]{I.~Okada}
\author[mpi]{I.~Popov}
\author[kit]{F.~Simon}
\author[utokyo]{X.~Shi}
\author[kek,sokendai]{S.~Tanaka}
\author[kek,sokendai]{K.~Uno}
\author[kek,sokendai,utokyo]{Y.~Ushiroda}
\author[nagoya]{B.~Urbschat}
\author[mpi]{H.~Windel}
\author[kek,sokendai]{R.~Zhang}

\address[uhm]{University of Hawaii, Honolulu, Hawaii 96822, USA}
\address[kmi]{Kobayashi-Maskawa Institute for the Origin of Particles and the Universe, Nagoya University, Nagoya 464-8602, Japan}
\address[kek]{High Energy Accelerator Research Organization (KEK), Tsukuba 305-0801, Japan}
\address[sokendai]{The Graduate University for Advanced Studies (SOKENDAI), Hayama 240-0193, Japan}
\address[nagoya]{Graduate School of Science, Nagoya University, Nagoya 464-8602, Japan}
\address[cincinnati]{University of Cincinnati, Cincinnati, Ohio 45221, USA}
\address[tmu]{Graduate School of Science, Tokyo Metropolitan University, Hachioji 192-0397, Japan}
\address[nara]{Nara Women’s University, Nara 630-8506, Japan}
\address[utokyo]{Department of Physics, University of Tokyo, Tokyo 113-0033, Japan}
\address[iowa]{Iowa State University, Ames, Iowa 50011, USA}
\address[mpi]{Max Planck Institute for Physics, 85748 Garching, Germany}
\address[kit]{Institute for Data Processing and Electronics, Karlsruhe Institute of Technology, 76131 Karlsruhe, Germany}

\begin{abstract}
The SuperKEKB/Belle II experiment aims to collect high-statistics data of B meson pairs to explore new physics beyond the Standard Model (SM). SuperKEKB, an upgraded version of the KEKB accelerator, has achieved a world-record luminosity of $4.71 \times 10^{34} \, \mathrm{cm^{-2}s^{-1}}$ in 2022 but continues to strive for higher luminosities. One of the major obstacles is Sudden Beam Loss (SBL) events, which cause substantial beam losses and damage to the Belle~II detector. To identify strategies for addressing SBL challenges, advanced beam diagnostic systems and enhanced beam abort systems have been developed. The diagnostic system aims to accurately pinpoint the start of beam losses, while the upgraded abort system quickly disposes of anomalous beams to minimize damage.

This paper details the development and implementation of these systems, including high-speed loss monitors, time synchronization with the White Rabbit system, and data acquisition systems. Efforts to understand the mechanisms of SBL events, using acoustic sensors to detect discharges, are also discussed. These measures aim to improve the operational stability and luminosity of SuperKEKB, contributing to the experiment's success.
\end{abstract}


\end{frontmatter}

\section{Introduction}
The SuperKEKB/Belle II experiment~\cite{Abe2010,Kou2019} aims to produce a large number of B meson pairs, equivalent to 50 ab$^{-1}$, to explore new physics beyond the Standard Model (SM) through quantum effects. By achieving high-statistics data collection, the experiment seeks to uncover new sources of CP violation and other phenomena not explained by the SM, potentially leading to the discovery of new physics.

SuperKEKB~\cite{Ohnishi2013}, the upgraded version of the KEKB accelerator~\cite{Toge1995,Kurokawa2003}, is designed to achieve a target luminosity of 6 $\times$ 10$^{35}$ cm$^{-2}$s$^{-1}$, a 30-fold increase over its predecessor. This significant enhancement is accomplished through comprehensive upgrades, including new beam optics, advanced beam monitors, upgraded RF systems, improved vacuum equipment~\cite{PhysRevAccelBeams.26.013201}, and enhanced magnets, including final focus superconducting magnets (QCS)~\cite{Ohuchi2022}. 

To handle the increased luminosity and higher beam backgrounds, it was necessary to upgrade the Belle detector at the interaction point (IP) to the Belle~II detector~\cite{Abe2010,Adachi2018}. The Belle~II detector utilizes several advanced subsystems, including the Vertex Detector (VXD), which consists of the innermost Pixel Detector (PXD) and the surrounding Silicon Vertex Detector (SVD), the Central Drift Chamber (CDC), the Time-Of-Propagation (TOP) detector, the Aerogel Ring Imaging Cherenkov (ARICH) detector, the Electromagnetic Calorimeter (ECL), and the KLong and Muon detector (KLM). These subsystems enable the Belle~II detector to provide precise measurements of particle trajectories and decay vertices, which are essential for studying rare decays and CP violation.

\begin{figure}[ht]
    \centering
    \includegraphics[width=\linewidth]{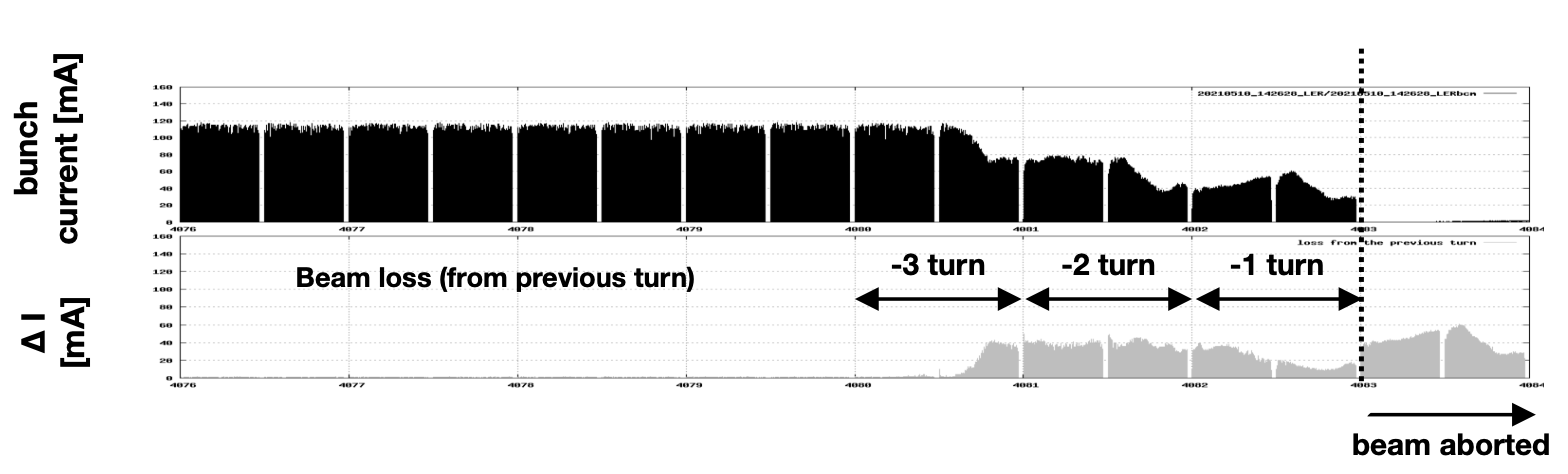}
    \caption{An example of an SBL event. The figure shows the bunch current (upper) and the change in bunch current from the previous beam-cycle (i.e., beam loss) (lower) for the seven beam-cycles before the beam abort. It is evident that the beam was lost over three beam-cycles before the abort. The post-abort beam loss indicates the amount of beam dumped due to the abort itself.}
    \label{fig:SBL}
\end{figure}

Currently, we are focused on collecting physics data while simultaneously working to improve the luminosity performance of SuperKEKB and maintaining the Belle~II detector. In June 2020, SuperKEKB surpassed the instantaneous luminosity record of KEKB, officially becoming a leading luminosity frontier machine. However, simultaneously, unexplained beam losses began to be observed. This beam loss, referred to as Sudden Beam Loss (SBL), is troublesome because it causes the beam to lose most of its particles within a few beam-cycles, scattering a large amount of radiation within the ring, as shown in Figure~\ref{fig:SBL}. The figure shows the bunch current measured by the Bunch Current Monitor (BCM)~\cite{BORBCM}.

This SBL phenomenon has caused damage to accelerator and detector equipment, such as the pixel detector in the innermost layers of the Belle~II detector and beam collimators. Since the radiation dose is increasing with the beam current, we have had to be cautious about increasing the current, which has hindered the improvement of luminosity.

By 2021, it became commonly recognized that this beam loss was not a transient issue but had some fundamental cause behind it, making it an unavoidable problem when increasing luminosity. Since then, we have started developing an advanced beam diagnostic system to localize the start of the beam loss, as well as enhancing the beam abort system to mitigate its effects. Although we are still striving to elucidate the cause of the SBL phenomenon, this article introduces some of the research efforts detailed in the sections on our beam diagnostic system and beam abort system upgrades, aimed at finding clues to address this issue.
\section{Overview of Key Accelerator Subsystems}
Before detailing the beam diagnostic system and beam abort system enhancements, which are central to addressing the SBL issue, we will briefly introduce the accelerator subsystems that are crucial to this study, particularly the collimator system and the beam abort system.

\subsection{Beam Backgrounds and Collimators}
Beam particles deviate from their normal trajectory due to scattering with residual gas in the ring~\cite{Nakayama2012}, interactions between particles within the bunch (Touschek effect~\cite{Piwinski1999}), beam injection oscillations, and other effects. These particles eventually collide with the beam pipe, creating secondary particle showers. To compensate for the inevitable loss of beam particles and to maintain the high beam currents necessary for achieving high luminosity, the machine operates in top-up mode, injecting new particles into the main ring at a rate of up to 25Hz per beam. Particle showers generated near the IP are detected as beam background~\cite{Natochii2023} in the Belle~II detector. Each subsystem has defined acceptable background levels, and shifters and experts closely monitor the background during operation. High background levels can increase trigger rates, destabilize the data acquisition (DAQ) system, and degrade the quality of the physics data.

\begin{figure}[ht]
\centering
\includegraphics[width=\linewidth]{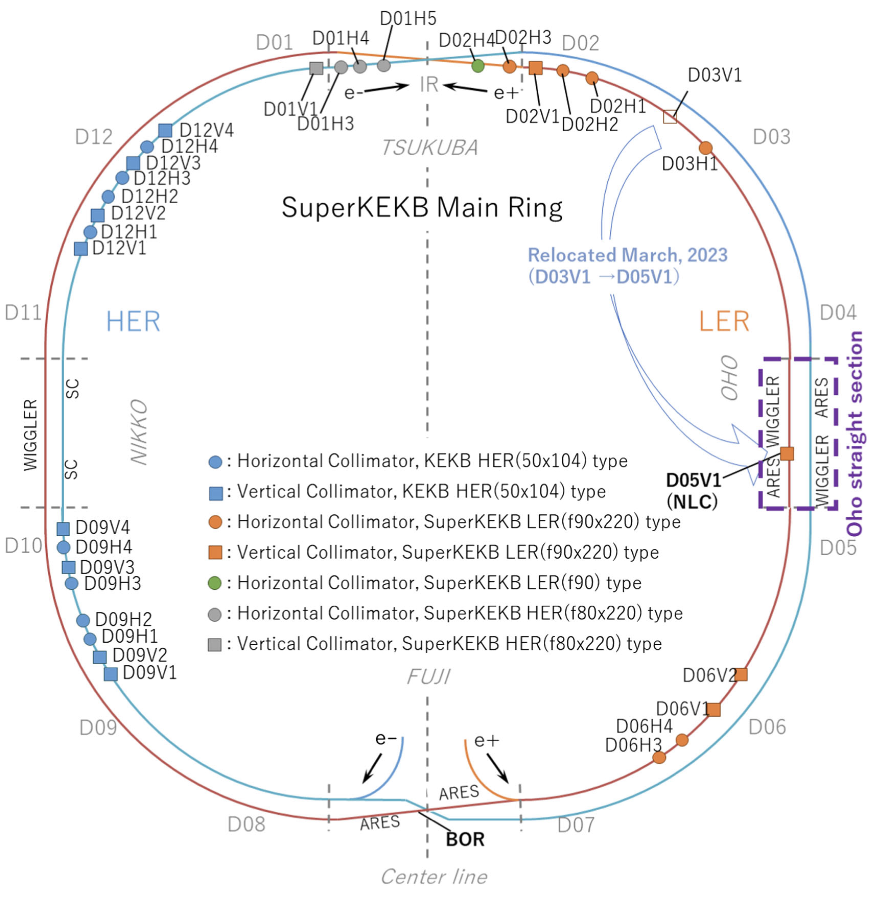}
\caption{Locations of the collimators in the SuperKEKB Main Ring. The diagram indicates the positions of horizontal and vertical collimators in both the High Energy Ring (HER) and Low Energy Ring (LER). The D05V1 collimator was relocated in 2023 and now includes the Nonlinear Collimator (NLC), is highlighted. This setup plays a crucial role in managing beam loss and minimizing background at the Belle~II detector and protecting the QCS final focus system.}
\label{fig:collimator_config}
\end{figure}

\begin{figure}[ht]
    \centering
    \includegraphics[width=\linewidth]{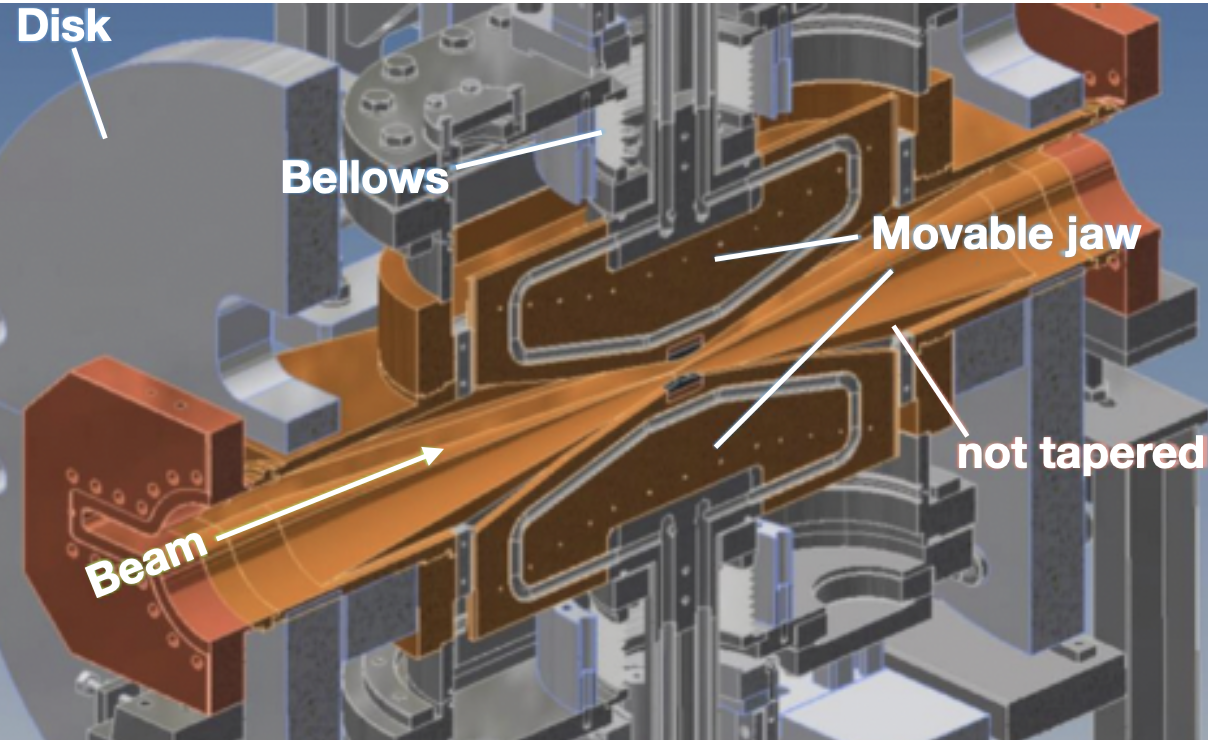}
    \caption{SuperKEKB-type vertical collimator. The movable parts (Jaws) on the vacuum side can move independently up and down.}
    \label{fig:vertical_collimator}
\end{figure}

Here, collimators~\cite{Ishibashi2020} come into play. SuperKEKB's ring is equipped with movable collimators, which are one of the accelerator components (see Figure~\ref{fig:vertical_collimator}). The tantalum head~\footnote{In addition to tungsten, other metals such as tungsten and titanium, as well as carbon, are also used for collimator heads. The length and material of the head in the beam direction are selected according to the collimator position.} attached to the movable part on the vacuum side is brought close to the beam orbit to scrape off as many deviated beam particles as possible before they reach the Belle~II detector. Collimators are also essential for protecting the QCS from quenching. There are 31 horizontal and vertical collimators installed upstream of the collision points in both the electron ring (HER) and positron ring (LER), as shown in Figure~\ref{fig:collimator_config}. 

Those used since the KEKB era are cantilevered collimators with heads on the top or bottom (inside or outside for horizontal collimators), but those newly installed for SuperKEKB~\cite{Ishibashi2020} are equipped with twin movable jaws, with independently operable jaws at the top and bottom. The apertures of each collimator are critical machine tuning parameters and must always be adjusted to optimal conditions. During operation, the apertures of each collimator are set to about 1 cm at the widest points and about 1~mm at the narrowest points, controlled with a precision of 50~$\mu$m.

Adjustments are frequently made amidst constantly changing operational conditions. Opening the collimators too much increases background levels, while closing them too much disrupts beam injection and increases the frequency of beam aborts. 

\begin{figure}[ht]
    \centering
    \includegraphics[width=\linewidth]{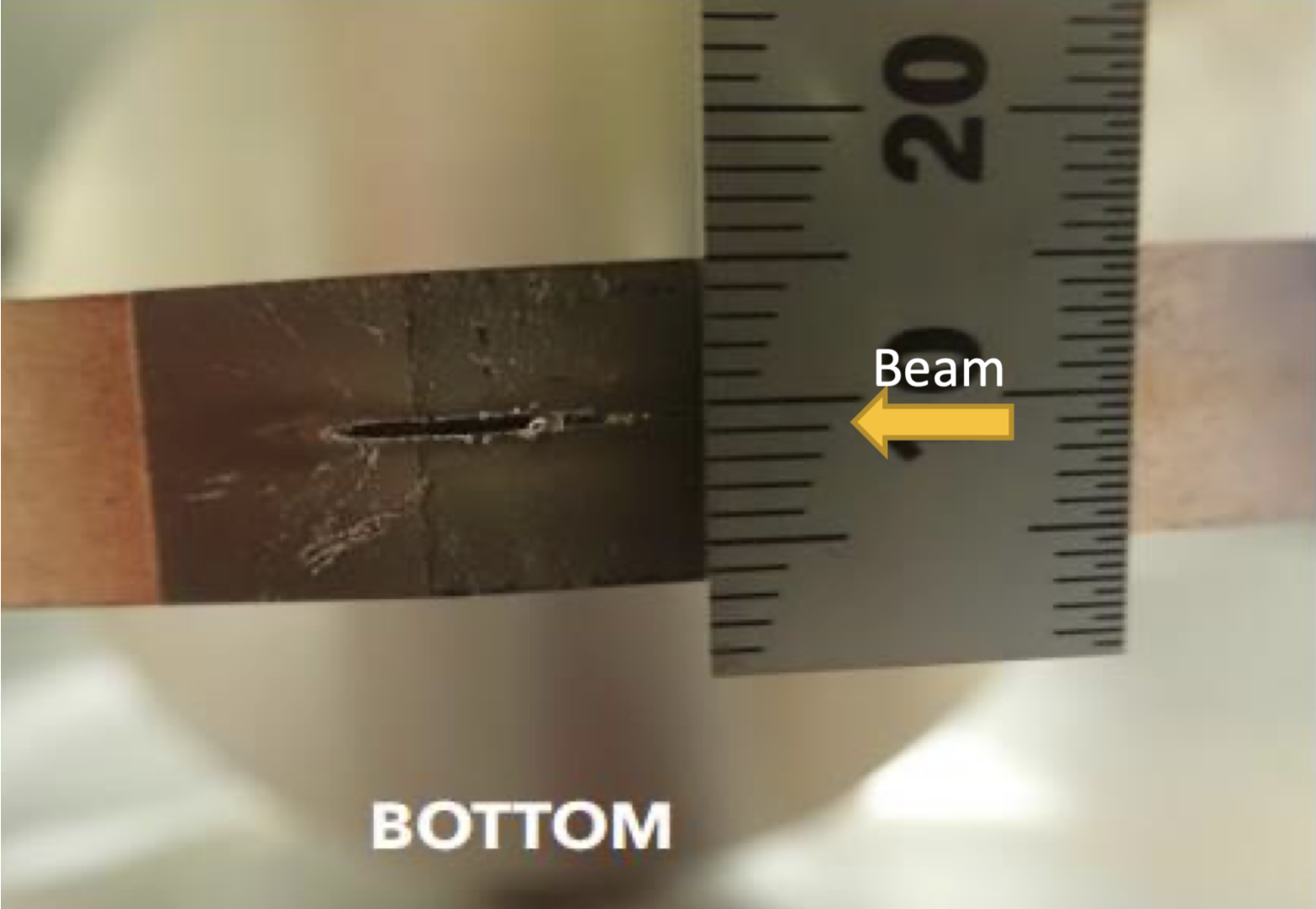}
    \caption{State of a damaged collimator head made of tantalum. Scratches can be seen along the beam axis direction~\cite{PhysRevAccelBeams.26.013201}.}
    \label{fig:collimator_damage}
\end{figure}

When a collimator is damaged by an SBL event (see Figure~\ref{fig:collimator_damage}), the aperture has to be widened, making background control difficult. Although other collimators can be used as substitutes, if key collimators close to the Belle~II detector, particularly vertical collimators, are damaged, there are no replacements. Depending on the extent of the damage, collimator heads may need to be replaced. This requires breaking the local vacuum, and while a full vacuum bakeout is no longer performed inside the tunnel, the reactivation of the NEG pumps is necessary. Restoring the ultra-high vacuum state typically takes about a day. Additionally, if the collimator is near the IR, beam backgrounds around the Belle~II detector temporarily increase due to collisions between residual gas and the beam. If a major collimator near the main ring injection point is damaged, the high radiation levels prevent immediate replacement. Waiting for radiation levels to decrease takes over a month, making replacement during the limited operation period impractical~\footnote{In a past incident involving damage to D06V1, high radiation levels required waiting for approximately one month after machine operation ended before the replacement could be safely conducted.}. Therefore, understanding the mechanism of the SBL phenomenon is essential to operate collimators without damage.

\subsubsection{Nonlinear Collimator (NLC)}
To mitigate the issue of collimator damage and improve beam stability, a novel collimation approach called the Nonlinear Collimator (NLC)~\cite{FausGolfe2006,Terui2023,Terui2024} was introduced during the Long Shutdown~1 (LS1) period, which spanned from the fall of 2022 to the spring of 2024. The NLC was installed in place of the wiggler magnets in the D05 section (see Figure~\ref{fig:collimator_config}). SuperKEKB is the first operational accelerator to adopt this NLC concept.

The NLC operates by using the nonlinear magnetic field of a skew sextupole magnet to kick beam halo particles with large vertical orbit deviations, removing them at a downstream vertical collimator. A second skew sextupole is then employed to return the remaining particles to their nominal orbit. The NLC has demonstrated promising results in mitigating beam halo without significantly affecting the beam core and reducing the overall ring impedance, thereby easing the conditions for transverse mode coupling instability. However, there is still room for improvement in mitigating injection background. This approach enhances the overall beam performance and stability, contributing to the reduction of beam background in the Belle~II detector. Furthermore, by minimizing the risk of collimator damage, the NLC ensures more reliable operation and reduces the downtime associated with collimator replacements. This innovative collimation method represents a significant advancement in maintaining the integrity of collimators and optimizing the performance of the SuperKEKB accelerator.


\subsection{Beam Abort System}
The beam abort system in SuperKEKB~\cite{Mimashi:IPAC2014-MOPRO023} is designed with both horizontal and vertical abort kicker magnets to safely and quickly dispose of the beam when anomalies are detected in the beam or accelerator equipment. The current system aggregates abort request signals from sensors installed around the ring via both optical fibers and electrical signal lines to the Central Control Building (CCB), which then transfers trigger signals to the abort kicker magnets. When the signal is received, the horizontal kicker magnet extracts the circulating beam, which is then deflected by the Lambertoson DC septum magnet towards the beam dump. The vertical kicker magnet protects the extraction window by tapering the beam trajectory, effectively enlarging the beam cross-section at the window. 

SuperKEKB's beams consist of two bunch trains, seprated by two abort gaps of approximately 200~ns each. This gap is set considering the rise time of the kicker magnets, and during an actual beam abort, trigger signals are sent at this abort gap timing to correctly apply the magnetic field to deflect the beam toward the beam dump. The time from anomaly detection to beam dump depends on the position of the sensor detecting the loss and the relationship between the lost bunch and the abort gap, but it is generally around 20 to 30~$\mu$s.

SuperKEKB uses various devices as abort sensors, including PIN photodiodes (PIN PD)~\cite{Ikeda2014}, ion chambers~\cite{Ikeda2014}, optical fibers, diamonds sensors~\cite{diamond_sensors,Bassi2021}, and CLAWS (plastic scintillators read out by SiPMs)~\cite{claws_sensor}. Below, we provide additional details on some of these sensors:

\subsubsection{PIN PDs and Optical Fibers}
PIN PDs~\cite{Ikeda2014} and optical fibers are installed in locations such as near collimators, where beam losses are likely to be observed because of their narrow aperture in the ring. They are used not only for beam aborts but also for real-time beam loss monitoring. These sensors are crucial for detecting and responding to SBLs.

\subsubsection{Ion Chambers}
Ion chambers~\cite{Ikeda2014} focus primarily on beam aborts, with each chamber covering several meters of the ring. They are widely installed to monitor radiation levels across the entire ring, ensuring comprehensive detection and contributing to the overall safety and functionality of the beam abort system.

\subsubsection{Diamond Sensors}
Diamond sensors~\cite{diamond_sensors,Bassi2021}, specifically single-crystal Chemical Vapor Deposition (sCVD) diamonds, are used as beam loss monitors at SuperKEKB. They are mounted on the beam pipe around the IP, SVD support cones, and QCS bellows. Diamond sensors function as solid-state ionization chambers. When high-energy particles pass through, they create electron-hole pairs, generating a current proportional to the radiation dose. This current is processed by dedicated electronics, including trans-impedance amplifiers and high-speed ADCs. The diamond sensors provide comprehensive coverage of the IP region. They deliver real-time data for monitoring and beam abort triggers, protecting the Belle~II detector from radiation damage and reducing the probability of QCS quenches. Their performance has been validated during the commissioning phases of SuperKEKB and Belle II, proving their reliability and effectiveness in beam loss monitoring.

\subsubsection{CLAWS Detectors}
The sCintillation Light And Waveform Sensors (CLAWS) system~\cite{claws_sensor} was introduced in 2016 during the SuperKEKB commissioning to monitor beam-induced backgrounds and has evolved in terms of layout and technology to adjust to changing requirements and use cases. Each CLAWS module utilizes plastic scintillators coupled with silicon photomultipliers (SiPMs) to detect and measure beam losses. The system comprises 32 modules, strategically placed around the IP Region, with 16 modules on the forward side and 16 on the backward side. These modules are positioned at four different longitudinal locations along the beam direction, approximately 1, 2, 3, and 4 meters from the IP, and at four different azimuthal angles (0$\degree$, 90$\degree$, 180$\degree$, and 270$\degree$). Initially designed for monitoring injection-induced backgrounds, CLAWS shifted its role to an abort sensor after 2020, as injection backgrounds decreased. Its rapid response compared to other sensors has proven invaluable in detecting beam losses early. CLAWS now provides real-time data on beam losses, helping to prevent potential damage and ensure stable operation, and its early detection capability has been effectively integrated into the current beam abort system~\cite{claws_commissioning_phase2, claws_proceedings, claws_scintillator}.

\section{High-speed Beam Diagnostic System}
To better understand beam losses, it is crucial to determine where in the ring the beam losses begin. The existing loss monitors, specifically PIN photodiodes, are designed to meet the requirements of the readout electronics, including accommodating a broad dynamic range. Consequently, these monitors exhibit prolonged integration and response times, limiting the measurement accuracy of beam loss timing to approximately 2-3~$\mu$s. This temporal resolution is inadequate for precisely identifying the initiation points of beam losses, thereby necessitating the deployment of high-speed beam loss monitors.

Loss monitors are strategically placed near collimators, as these locations represent the smallest physical apertures in the ring and are thus more prone to beam losses. Therefore, both the current PIN PDs and the newly developed loss monitors are predominantly installed in proximity to the collimators. Given that the collimator intervals range from a few tens of meters to several hundred meters, a time resolution of approximately 30~ns for these high-speed loss monitors is sufficient, obviating the need for exceptionally fast detectors.

\begin{figure}[ht]
    \centering
    \includegraphics[width=78mm]{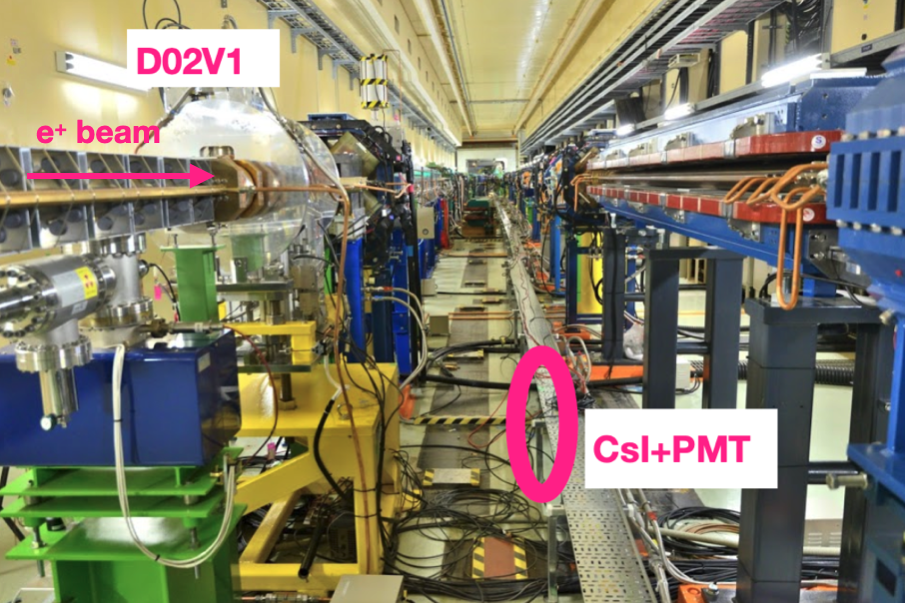}
    \includegraphics[width=78mm]{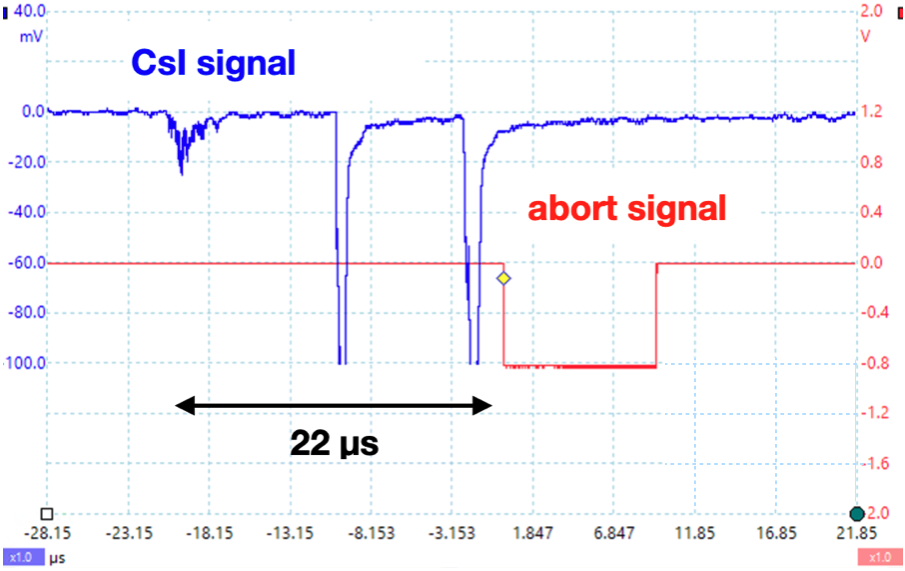}    
    \caption{Loss monitor installed near the vertical collimator (D02V1) in LER (upper photo) and signal observed with the loss monitor (lower figure, blue line). Loss is observed over three turns.}
    \label{fig:D02V1}
\end{figure}

To demonstrate the rapid detection of beam loss, a PMT coupled with a CsI crystal was installed near the vertical collimator in LER, which is the closest to the Belle~II detector. This setup successfully recorded clear beam loss signals (see Figure\ref{fig:D02V1}). Analysis revealed that beam losses started two beam-cycles~\footnote{One beam-cycle corresponds to approximately 10~$\mu$s.} before the abort signal was issued by the CLAWS sensors at the IP. By increasing the number of loss monitors and conducting synchronized measurements across the monitors, it will be possible to determine the initial detection point of the beam loss. This integration of multiple loss monitors with a time synchronization system constitutes the high-speed beam diagnostic system.

\subsection{High-speed Loss Monitors}\label{sec:LM_configuration}
\begin{figure}[htb]
\begin{center}
\includegraphics[width=78mm]{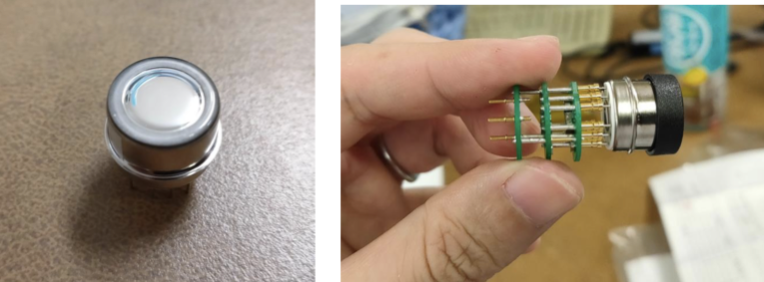}
\end{center}
\vspace{-5mm}
\caption{EMT (R15435-01). When installed, it is shielded in an aluminum box.}
\label{fig:EMT}
\end{figure}

In addition to the PMT with a CsI crystal introduced earlier, an Electron Multiplier Tube (EMT) was adopted as a loss monitor (see Figure~\ref{fig:EMT}). The former has a time constant of about 20~ns, meeting our required time accuracy. The latter was originally developed and considered for use as a muon beam monitor in the T2K experiment~\cite{EMT_paper}. Although the dynode structure is the same as that of a general PMT, the photocathode is replaced with an aluminum film, making it less sensitive to photons but more radiation-resistant than commonly used SiPMs.

\begin{figure}[htb]
\begin{center}
\includegraphics[width=78mm]{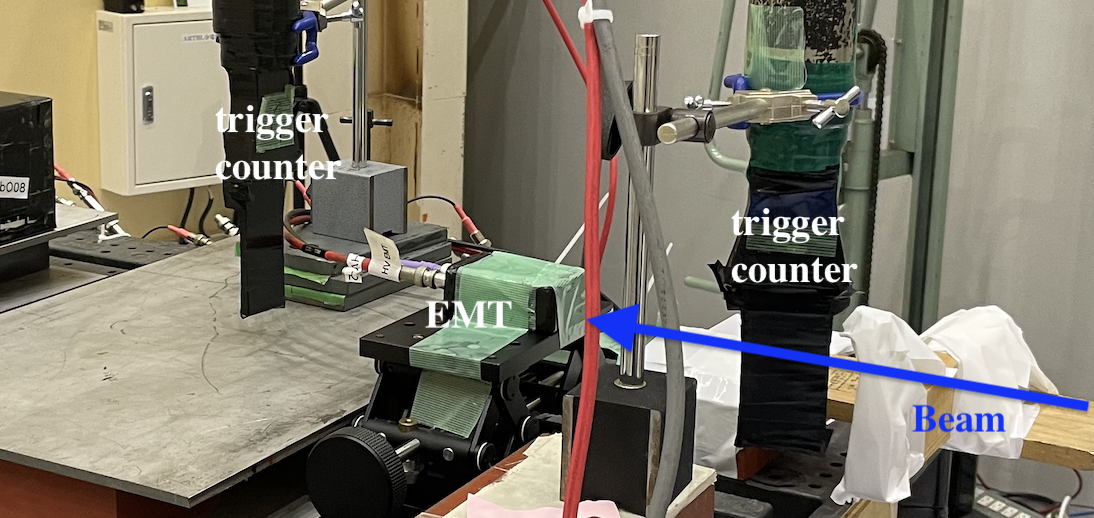}
\end{center}
\vspace{-5mm}
\caption{Setup of the EMT test using the 4~GeV electron beam at the newly established test beam line at KEK PF-AR facility.}
\label{fig:EMT_beamtest}
\end{figure}

The performance of EMT was evaluated using a Sr-90 beta source and electron beams at the newly established test beam line at KEK’s PF-AR facility. During the beam tests, triggers were created with counters placed before and after the EMT, as shown in Figure\ref{fig:EMT_beamtest}, and basic characteristics such as detection efficiency were evaluated.

The EMT signal pulse width was confirmed to be approximately 10~ns, meeting the required precision. The detection efficiency was measured as a function of electron beam energy within the range of 1-5~GeV and found to be around 0.5\% per electron. This efficiency indicates that for a bunch current of 0.5~mA (corresponding to approximately $3.1\times 10^{10}$ electrons) with an acceptance of 0.1\%, where 0.1\% of the shower particles enter the EMT's fiducial volume, the system is capable of detecting the loss of electrons from a single bunch. 

\subsection{System Configuration}
\begin{figure}[ht]
\centering
\includegraphics[width=\linewidth]{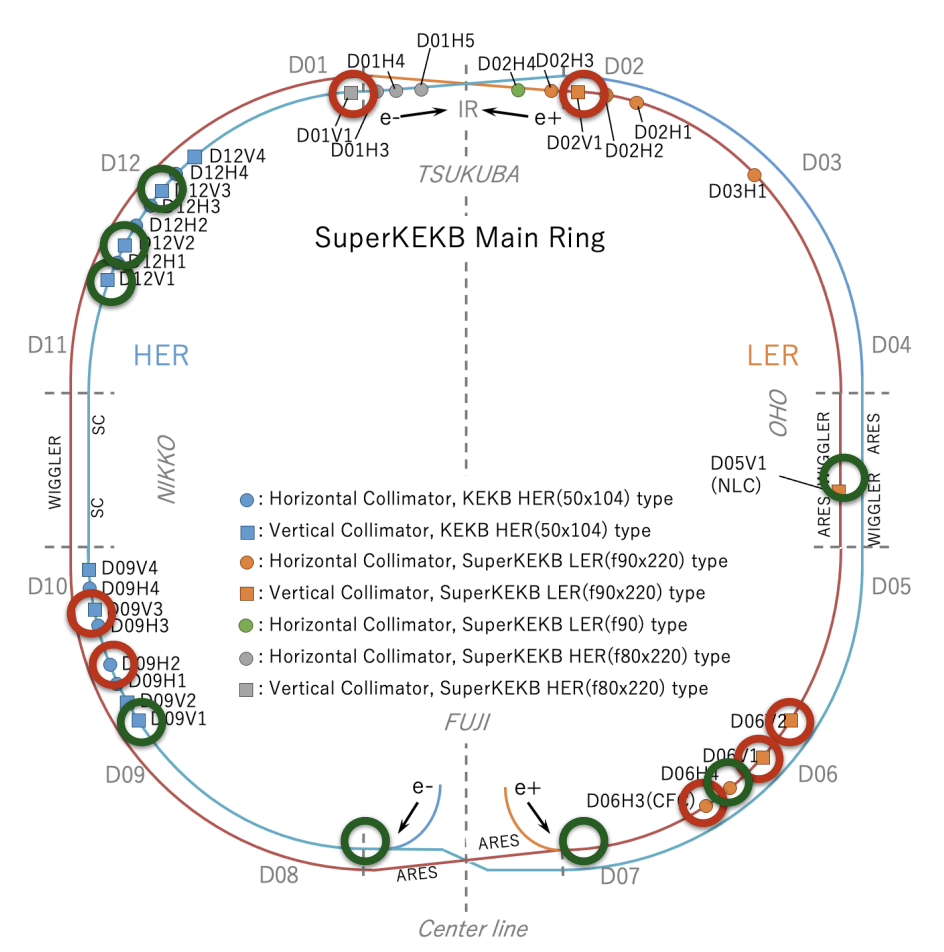}
\caption{Green circle indicates the loss monitors installed during Long Shutdown 1 (LS1), and maroon circle shows those installed before the LS1 period. The 15 locations (7 in LER and 8 in HER) cover the entire ring, enabling precise identification of beam loss points.}
\label{fig:LM_locations}
\end{figure}

During machine operation, the entire tunnel experiences high radiation levels, especially the part of the ring near the beam injection point, due to significant beam loss associated with the injection process. Therefore, EMTs, which possess high radiation tolerance, were installed in the upstream section near the injection point, while CsI+PMTs were installed downstream near the Belle~II detector. Due to the potential lack of radiation tolerance of the data acquisition system, it was placed in the nearest power station, with long signal and power cables, approximately 100-300~m in length, running through the tunnel. To date, 15 loss monitors have been installed primarily around the key collimators in both the LER and HER rings, as shown in Figure~\ref{fig:LM_locations} and detailed in Table~\ref{tab:loss_monitor_locations}.

\begin{table}[ht]
\centering
\caption{Locations of Installed Loss Monitors (LM)}
\vspace{1mm}
\begin{tabular}{|c|c|p{0.25\textwidth}|}
\hline
\textbf{LM Type} & \textbf{Ring} & \textbf{Position} \\
\hline
EMT & LER & D06H3, D06H4, \newline Injection Point \\
\hline
CsI+PMT & LER & D06V1, D06V2, \newline D05V1, D02V1 \\
\hline
EMT & HER & D09H2, D09V1, \newline D09V3, Injection Point \\
\hline
CsI+PMT & HER & D12V1, D12V2, \newline D12V3, D01V1 \\
\hline
\end{tabular}
\label{tab:loss_monitor_locations}
\end{table}

\subsection{Time Synchronization System}
\begin{figure}[htb]
\begin{center}
\includegraphics[width=78mm]{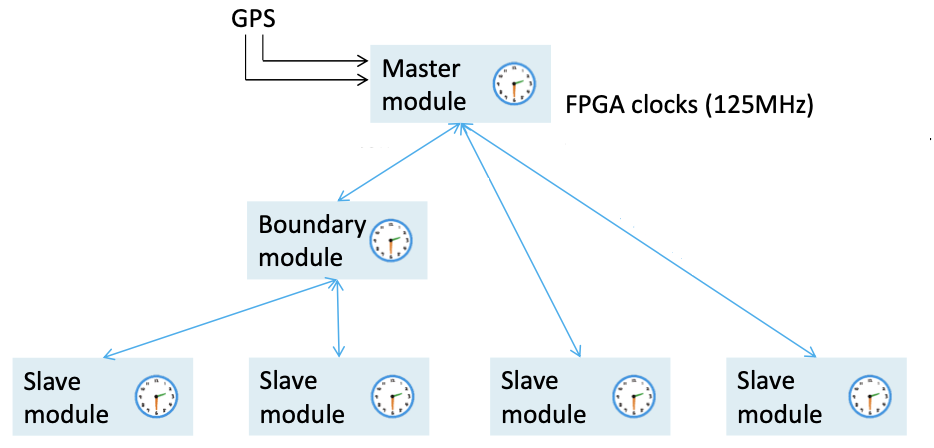}
\end{center}
\vspace{-5mm}
\caption{Conceptual diagram of the WR system. The boundary module relays time synchronization from the master module to slave modules, allowing the system to extend beyond the 80~km limit of a single fiber link.}
\label{fig:WR}
\end{figure}

To compare the relative time differences between loss monitors, we adopted the White Rabbit (WR) high-precision time synchronization system~\cite{WhiteRabbit_ref}. WR achieves sub-nanosecond time synchronization by synchronizing the internal clocks (e.g., CPU time) of different WR nodes connected via optical cables~\footnote{Direct attach cables connected to SFPs are also acceptable.} in three stages: sub-microsecond time synchronization via the Precision Time Protocol (PTP), internal clock synchronization via Synchronous Ethernet (Sync-E), and identifying and correcting phase differences between synchronized clocks via digital dual mixer time difference. Originally developed primarily by CERN, the WR standard is fully open-source on the Open Hardware Repository (OHR), and its implementation is progressing in various physics experiments. WR nodes are categorized as either Master or Slave, connected by single-mode optical fibers to share timestamps. In cases where distances exceed 80~km, the WR master module can be operated as a boundary module, facilitating daisy-chain connections that extend time synchronization, as depicted in Figure~\ref{fig:WR}. While SuperKEKB’s 3~km ring circumference does not necessitate daisy-chaining, this capability highlights the scalability of the WR system for larger accelerator facilities. The WR-TDC~\footnote{WR can also be used as a TDC or ADC, depending on the FMC card installed.} we utilize operates with a 125~MHz clock, achieving a time resolution of 8~ns, which sufficiently meets our precision requirements.

\subsection{Data Acquisition System}
\begin{figure}[htb]
\begin{center}
\includegraphics[width=78mm]{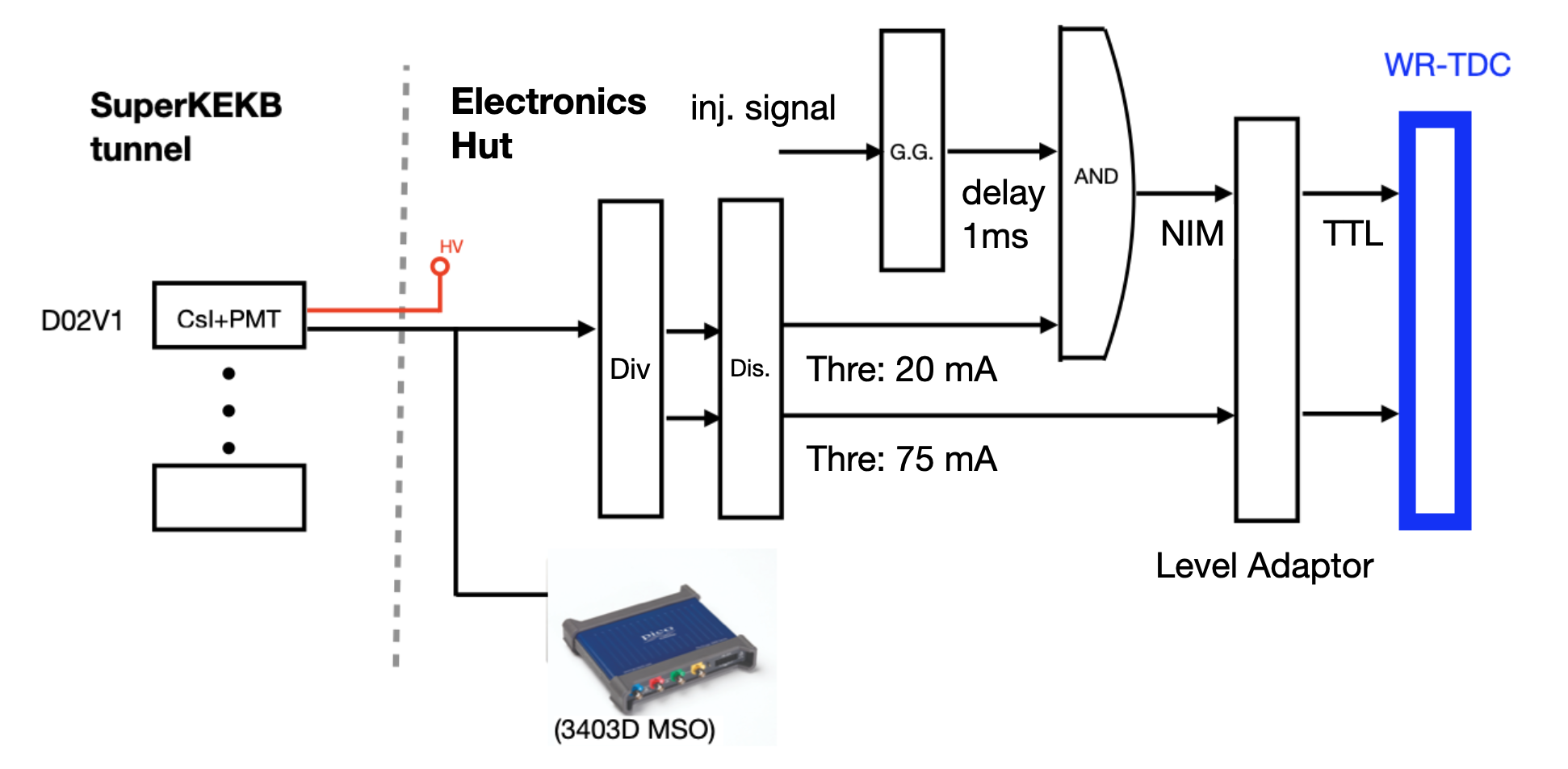}
\end{center}
\vspace{-5mm}
\caption{Schematic of the loss monitor data acquisition system.}
\label{fig:DAQ}
\end{figure}

In addition to WR-TDC, we use a USB oscilloscope (Picoscope 3403D) to acquire data from the loss monitors. These two systems operate independently, with the oscilloscope capturing waveform data 1~ms before and after the abort signal (corresponding to approximately 100 beam-cycles before and after), enabling detailed time-series analysis of beam losses. As shown in Figure~\ref{fig:DAQ}, loss monitor signals are split into two at the nearest power station, with one path connected to the oscilloscope and the other to the WR-TDC.

The loss monitors can observe not only SBL events but also injection losses arising from the horizontal beam oscillations during the beam injection. Therefore, the signal line on the WR side is further split to obtain data for beam injection studies, in addition to SBL events. For SBL-targeted signals, the threshold is set to the minimum, and an injection veto is applied to the WR-TDC until the injection oscillations settle.

\section{Beam Loss Analysis}

\subsection{Timing Calibration for SBL Analysis}
To accurately analyze beam losses and SBL events, precise timing calibration of the detection system is crucial. The WR system provides accurate time synchronization by sharing timestamps between nodes, eliminating the need for special corrections. The WR-TDC and Picoscope timing were further calibrated using cable length corrections. These corrections account for the cable length between the loss monitors and the WR slave nodes at each station. The cable lengths were measured using a network analyzer by observing the signal reflection peaks at connector positions.

For SBL analysis, the beam abort signal is used as the time reference. The master WR node, located in the central control building (CCB), is directly connected to the abort module and thus inherently knows the exact time the abort signal is issued. However, for the Picoscope at each power station, corrections are necessary to account for the time delay caused by the signal propagation from the CCB to each station where the loss monitor DAQ systems are installed. This ensures accurate determination of which loss monitor first detected the beam loss by comparing the timestamps from the different monitors.

\subsection{Injection Analysis}
It is essential to maintain or improve beam injection performance as we aim for higher luminosity. Higher beam currents require increased beam injection, and more frequent injections are necessary when the beam is squeezed due to its shorter lifetime. This is particularly critical in HER, where injection efficiency is lower compared to LER. The high-speed loss monitors play a vital role in achieving this objective by providing real-time data that helps in optimizing beam injection performance. Additionally, the loss rate is displayed live in the accelerator control room, assisting in diagnosing issues with the collimators.

\begin{figure}[htb]
\begin{center}
\includegraphics[width=78mm]{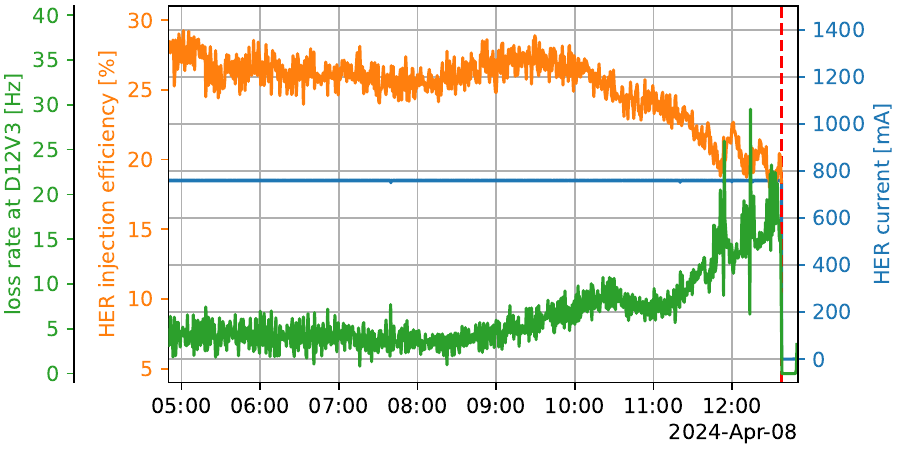}
\end{center}
\vspace{-5mm}
\caption{Correlation between beam loss rate (green) observed by the high-speed loss monitors near the D12V3 collimator in HER and the injection efficiency (orange) of HER when the current (blue) in HER was 760 mA.}
\label{fig:injection_monitor}
\end{figure}

We analyzed the beam losses occurring during beam injection. Figure~\ref{fig:injection_monitor} illustrates the correlation between the beam loss rate observed by the high-speed loss monitor installed near one of the HER collimators and the injection efficiency of HER under constant beam current conditions. The data shows that the loss rate increases as the injection efficiency decreases. This correlation is continuously monitored across all loss monitors, aiding in the tuning of injection parameters.

\subsection{SBL Analysis}

\subsubsection{Pre-LS1 (2022 Analysis)}
\begin{figure}[htb]
\begin{center}
\includegraphics[width=78mm]{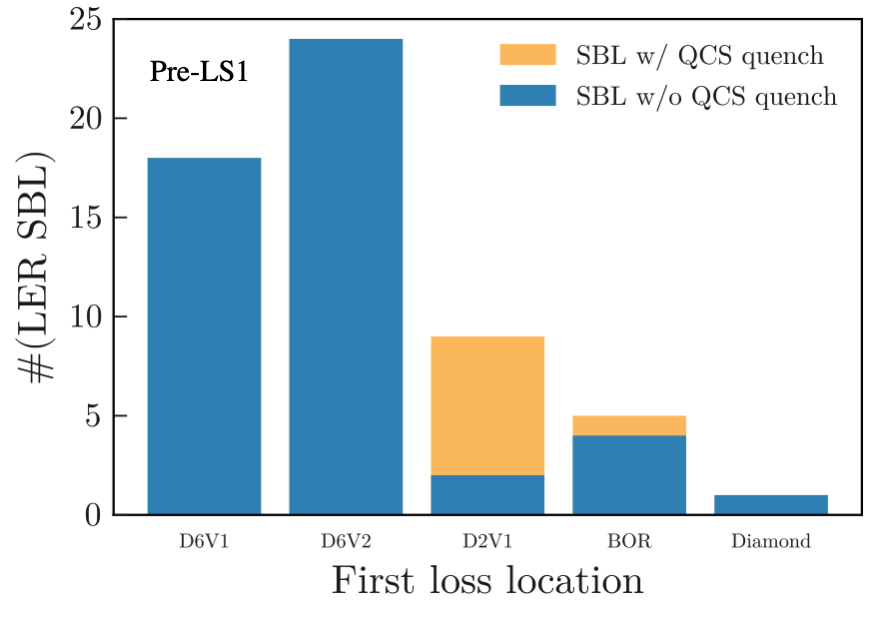}
\includegraphics[width=78mm]{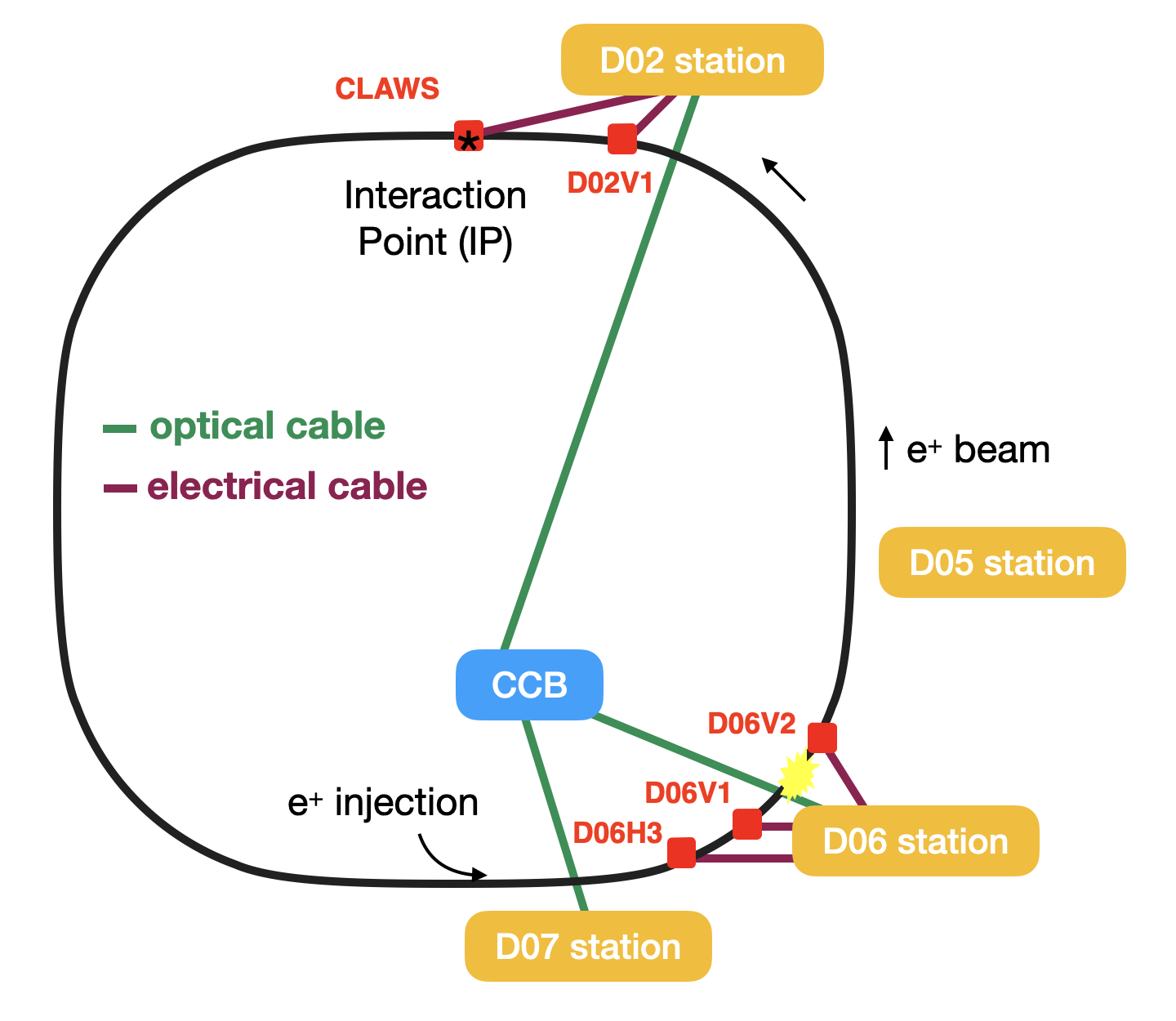}
\end{center}
\vspace{-5mm}
\caption{Summary of SBL events during the 2022 machine operation (upper figure) and the positional relationship between key collimators in LER (square marks) and the accelerator ring. The yellow mark, placed between D06V1 and D06V2, highlights a general trend where initial beam losses are often detected. Note: The optical cable path shown running along the inner side of the ring and the electrical cable path shown running along the outer side of the ring are for illustration purposes only; in reality, both pass through the tunnel (lower figure).}
\label{fig:SBL_stat}
\end{figure}

We analyzed the SBL events observed during the 2022 operation. Although SBL events occur in both LER and HER, they are more frequent in LER, so the analysis focused on SBL events in LER. Using the beam abort signal timing~\footnote{This is the timing when the abort trigger module in the CCB receives an abort request from the abort sensor and issues an abort signal.} as a reference, we investigated which loss monitor first detected the beam loss. It was found that beam losses were often first detected at D06V1 or D06V2, located upstream in the ring and closer to the injection point than the collision point, as shown in Figure~\ref{fig:SBL_stat}.

It was also found that when the first loss was detected at D02V1, which is close to the Belle~II detector, the probability of a QCS quench was high. It is conceivable that the beam trajectory was disturbed upstream of D06V1, resulting in beam loss near the D06V1 collimator, and the generated shower was observed by the loss monitor. The increased number of SBL events at D06V2 was likely due to damage to D06V1 during machine operation, leading to the widening of the collimator aperture and thus increasing the number of events passing through D06V1 and hitting D06V2. Additionally, despite beam losses being observed upstream, the first abort request was often issued not by the abort sensors placed at D06V1 or D06V2, but by CLAWS placed near the Belle~II detector~\footnote{Abort requests are issued from various locations around the ring, but the first signal to reach the CCB and trigger the abort is dominantly from CLAWS.}.

\subsubsection{Post-LS1 (2024 Analysis)}
\begin{figure}[htb]
\begin{center}
\includegraphics[width=78mm]{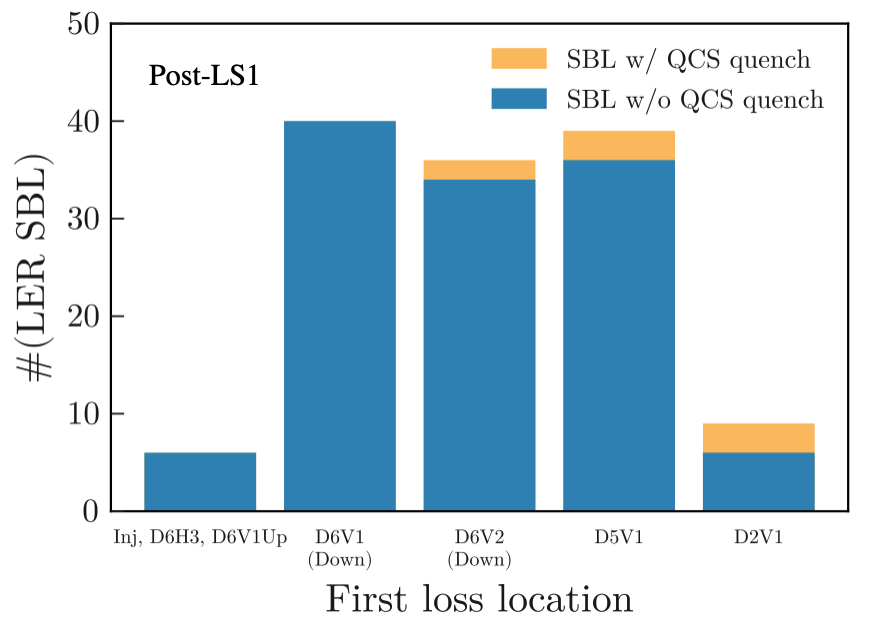}
\end{center}
\vspace{-5mm}
\caption{Summary of SBL events during the 2024 machine operation. For the collimator and the loss monitor locations are illustrated in Figure~\ref{fig:LM_locations}.}
\label{fig:SBL_stat2024}
\end{figure}

During the LS1 period, three loss monitors were added to LER and five to HER (see Figure~\ref{fig:LM_locations}), allowing for extended beam diagnostics. This significant enhancement has enabled more precise monitoring and analysis of beam losses. In the 2024 analysis, 132 SBL events were observed in the LER and 19 in the HER, providing a substantial dataset for investigation (see Figure~\ref{fig:SBL_stat2024}). The frequency of SBL events in the LER was approximately three times higher than in 2022. Initially, beam losses were frequently detected first at D06V2 and D05V1, where the NLC collimator was installed during LS1. Over time, there was a noticeable shift, with an increasing number of losses being detected first at D06V1. This shift is likely influenced by the aperture settings of each collimator.

\begin{figure}[htb]
\begin{center}
\includegraphics[width=78mm]{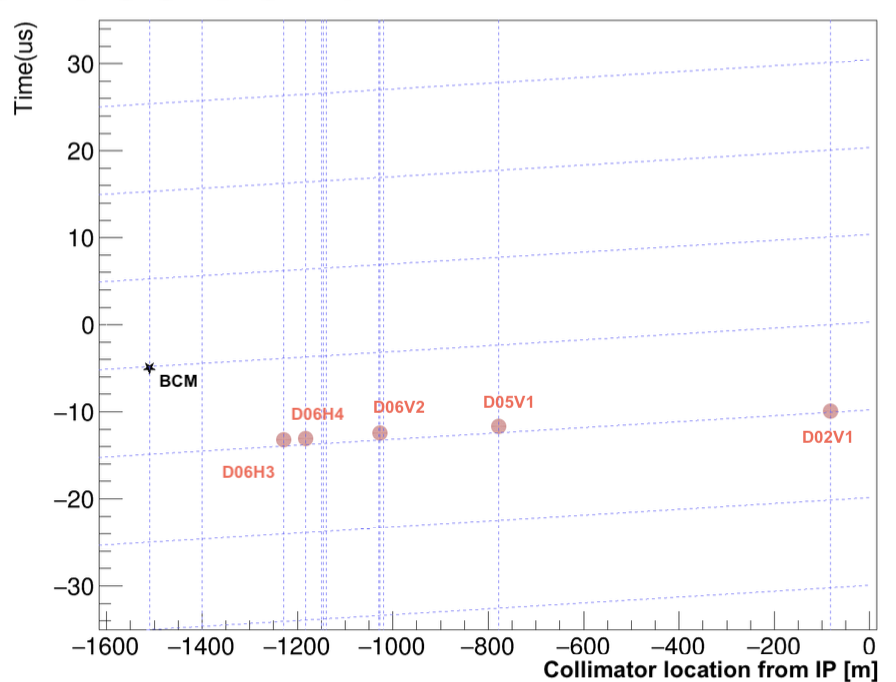}
\end{center}
\vspace{-5mm}
\caption{Beam losses detected by various LER loss monitors during an injection kicker misfire. The horizontally kicked bunches collide with the D06H3 collimator, generating a shower that is subsequently detected by the downstream loss monitors. The dashed diagonal lines represent the natural delay in beam arrival time across different collimator locations, due to the beam's circular trajectory around the ring. For example, the distance from D06H3 to D02V1 is approximately 1150~m, which introduces a beam arrival delay of around 3.8~$\mu$s.}
\label{fig:inj_kicker_misfire}
\end{figure}

SBL events caused by the misfire of a thyratron switch in the injection kicker magnet were also analyzed. In these cases, losses were predominantly first detected at D06H3. This can be explained by the horizontally kicked bunches colliding with the D06H3 collimator, with the resulting shower observed by the loss monitor located just downstream of the collimator, as shown in Figure~\ref{fig:inj_kicker_misfire}. The figure shows that the loss times and positions at each loss monitor are well-aligned in a two-dimensional plane, confirming that the time calibration is functioning correctly.

Additionally, there is a notable correlation between SBL frequency and beam current; higher beam currents are associated with more frequent SBL events. Furthermore, a large number of LER SBL events were found to correlate with pressure bursts exceeding $1 \times 10^{-7}$ Pa, with relative pressure changes surpassing 20\% per second, at specific locations within the ring, particularly downstream of the D10 and D04 sections. The first loss location strongly correlates with the IR dose. For SBLs with the first loss occurring at D02V1, the average IR dose is 849~mRad. In contrast, for SBLs with the first loss occurring at D06V1, the average IR dose is only 45~mRad. This difference is reasonable as the collimators between D06V1 and the IR can mitigate the loss in the IR region.

In the HER, SBL events were primarily first detected by loss monitors installed in the D09 section, such as D09V1 and D09V3. However, no significant correlation has been observed between these losses and pressure bursts or other potential indicators. Continued investigation and analyses are crucial to better understand the mechanisms driving these SBL events in the HER.

\subsection{Mechanism of SBL}
Despite gaining insights into the locations of beam losses through the beam diagnostic system, the fundamental causes of SBL events remain elusive. Previously, similar unexpected beam losses have been observed and reported in other accelerators. For instance, at SLAC’s PEP-II, rapid beam losses due to vacuum breakdown were eventually identified as being caused by damaged tiles that absorbed higher-order harmonics excited by passing bunches, following an extensive investigation~\cite{PEP-II-tile_damage}. Similarly, at CERN’s LHC, beam losses attributed to Unidentified Falling Objects (UFOs) are currently under investigation~\cite{LHC_UFO}.

At SuperKEKB, several hypotheses have been proposed to explain the occurrence of SBL events. One prominent theory involves beam-dust interactions, where dust particles generated from damaged components such as clearing electrodes fall into the beam path, leading to interactions that result in beam losses~\cite{PhysRevAccelBeams.26.013201,Suetsugu2010,Ikeda2024}. 
This hypothesis is supported by the observed correlation between SBL events and pressure bursts in specific sections of the accelerator, particularly those exposed to synchrotron radiation.

Another hypothesis is the fireball hypothesis, which suggests that high-temperature microparticles with a high sublimation point, potentially originating from damaged collimators, might become heated by the beam’s electromagnetic fields~\cite{fireball}. These particles could float within the beam pipe and, upon striking the pipe’s surface with lower sublimation point, generate plasma. As the plasma grows, it may absorb microwaves and ultimately cause a significant vacuum discharge. This discharge could then interact with the beam, leading to rapid losses. 

\section{Acoustic sensors}
Regardless of whether the cause of SBL is beam-dust interaction or fireballs, the generated microparticles interact with the beam, causing various effects that may include discharges. Detecting these phenomena would provide conclusive evidence for the cause of SBL. One effective method to observe such interactions is through acoustic emission (AE) sensors.


Acoustic sensors were introduced based on the knowledge that discharges in RF cavities can be identified by ultrasonic wave signals~\cite{Frisch:2000wx}. These AE sensors measure sound waves generated by the thermal shock of discharge currents impacting metal surfaces, in the ultrasonic range to avoid environmental noise. Even if discharges do not occur, particles or remnants from beam-dust interactions may still collide with the beam pipe, generating thermal shocks or mechanical vibrations that can be detected by the acoustic sensors. Acting as ``new ears" to complement the ``new eyes" of high-speed loss monitors, AE sensors are valuable for detecting both discharges and beam-dust interactions that may be related to SBL events.

\begin{figure}[htb]
\begin{center}
\includegraphics[width=78mm]{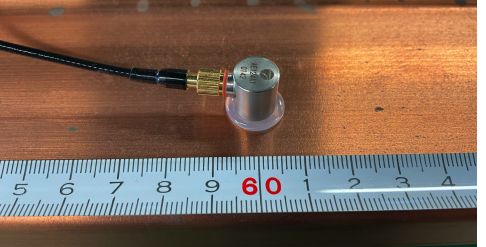}
\end{center}
\vspace{-5mm}
\caption{Acoustic sensor (AE124AT) considered for installation around collimators.}
\label{fig:acoustic_sensor}
\end{figure}

AE sensors (Model AE124-AT), designed by MATEX KENZAI CORPORATION, were first evaluated for performance and adhesive suitability, followed by the optimization of sensor positions and numbers using spare collimators. These sensors detect sound waves through a piezoelectric element, which generates a voltage when subjected to inertial forces, making them highly sensitive to mechanical vibrations. Observations of acoustic signals during vacuum discharges in RF cavities were also conducted. A total of 34 AE sensors, each measuring 1~cm $\times$ 1~cm, were then installed at key locations in the SuperKEKB tunnel, including the collimators at D02V1, D05V1, D06V1, D06V2, and QCS-R~\footnote{An AE sensor was installed near QCS-R during LS1 after discharge marks were discovered in this area when the QCS was opened.}. These sensors, which measure vibrations in the high-frequency range (10~kHz to 1~MHz), are particularly suited for this purpose. Figure~\ref{fig:acoustic_sensor} shows the setup of one such AE sensor.

\begin{figure}[htb]
\begin{center}
\includegraphics[width=78mm]{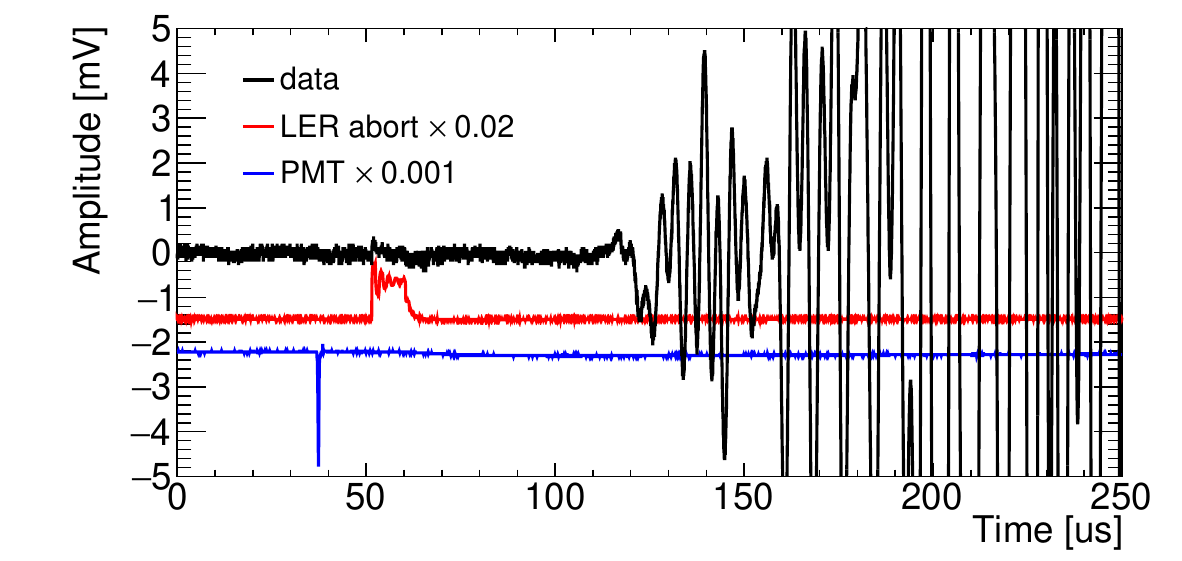}
\end{center}
\vspace{-5mm}
\caption{Typical signal measured by the AE sensor, with the readout oscilloscope triggered by the beam abort signal. The black line represents the acoustic signal data, the blue line represents the loss monitor signal, and the red line represents the beam abort trigger signal. The acoustic signal is observed after the beam loss is detected.}
\label{fig:AE_sensor_signal}
\end{figure}

Sound waves have been measured with the AE sensors since the start of operations in 2024. A typical AE sensor signal is shown in Figure~\ref{fig:AE_sensor_signal}, where the readout oscilloscope is triggered by the beam abort signal. This indicates that the ultrasonic wave signal is detected when the beam loss occurs, and there is a relevant electromagnetic shower passage at the accelerator component equipped with AE sensors, confirming the basic functionality of this method. Near the acoustic sensors, a high-speed loss monitor explained in Section~\ref{sec:LM_configuration} was installed. 

If a discharge inside a vacuum chamber causes beam loss, the AE signal would be expected to precede the loss monitor signal. However, after accounting for the sound wave propagation time of approximately 80~$\mu$s~\footnote{The actual propagation time is determined for each event by fitting the waveforms with a resonator model.}, the AE signal has consistently been observed to arrive later than the loss monitor signal. Therefore, the fireball hypothesis is disfavored as the cause of SBL, but its possibility cannot be entirely ruled out. Further investigation near upstream collimators is necessary. On the other hand, the AE sensor is established as a new method to detect beam loss. This allows inactive materials such as vacuum chambers and collimators to be turned into active devices for detecting beam loss.

\section{Beam Abort Upgrade}
In the development of the loss monitor system and during the operation of SuperKEKB, it became evident that speeding up the beam abort to minimize damage caused by beam losses is as important as understanding SBL. As shown in Figure~\ref{fig:SBL_stat}, the D06 power station is geographically closer to the central control building (CCB) and abort kicker magnets than the collision point where the CLAWS system was initially installed, near the interaction region (IR). However, the first abort request was issued by CLAWS. This observation led to the crucial insight that if the first abort request could be issued from sensors placed at D06 instead of CLAWS in IR, the abort could be significantly sped up.

To verify this effect, we used the loss monitor installed at D06V1 to output a pseudo-abort request and compared it with the abort request issued by CLAWS during the last period of beam operation in 2022. We demonstrated that an improvement of up to 10~$\mu$s was possible. An improvement of 10~$\mu$s corresponds to approximately one beam-cycle, which would reduce beam losses in the ring by about 20\% in the example shown in Figure~\ref{fig:SBL}. For some SBL events, even greater improvements can be expected. Based on this promising result, we decided to install additional CLAWS in the upstream regions of the ring, 
specifically around the vertical collimators in the D05 and D06 areas.

\begin{figure}[htb]
\begin{center}
\includegraphics[width=78mm]{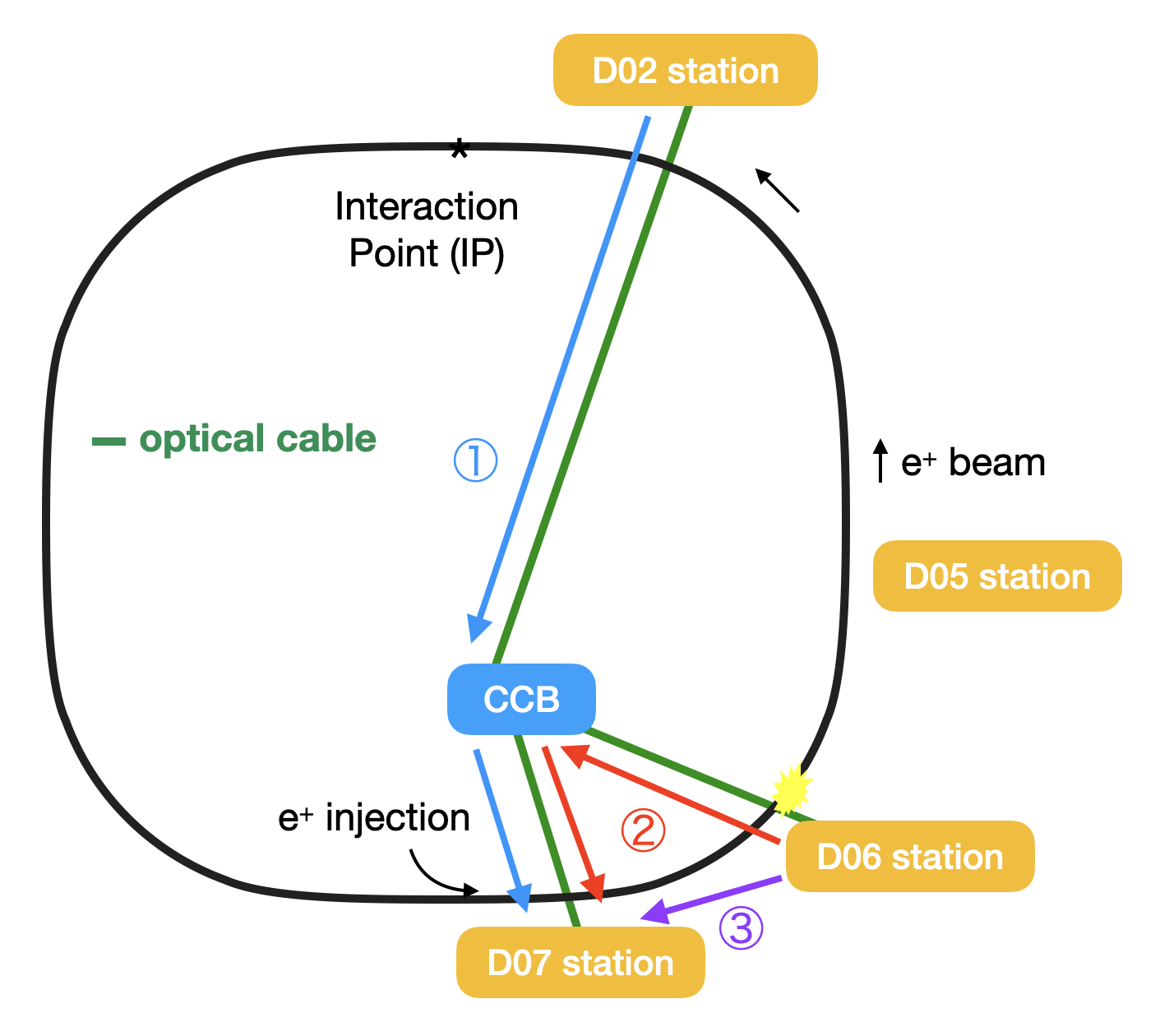}
\end{center}
\vspace{-5mm}
\caption{Ideas for abort speedup. The arrows indicate the current path of abort requests from CLAWS (1), the path if the abort request could be issued from D06 (2), and the path bypassing CCB (3). The yellow mark indicates the location of the first beam loss. Note: The optical cable path is shown running along the inner side of the ring for illustration purposes; in reality, it passes through the tunnel. While the diagram focuses on D06, the same discussion applies to D05, although the D05 cable path is not shown in the figure.}
\label{fig:fast_abort_scenarios}
\end{figure}

Furthermore, as shown in Figure~\ref{fig:fast_abort_scenarios}, if the abort request signal can be sent directly from the D06 power station to the D07 power station where the abort kicker pulse power supply is located, bypassing the CCB, even greater improvements can be expected. Additionally, replacing optical fiber communication with laser communication for signal transmission between power stations would enable further speedups. Currently, we are working on expanding the abort modules to the D07 power station and developing laser transport R\&D with this ambitious goal in mind.

\subsection{Additional CLAWS Installation at D05}
\begin{figure}[htb]
\centering
\includegraphics[width=\linewidth]{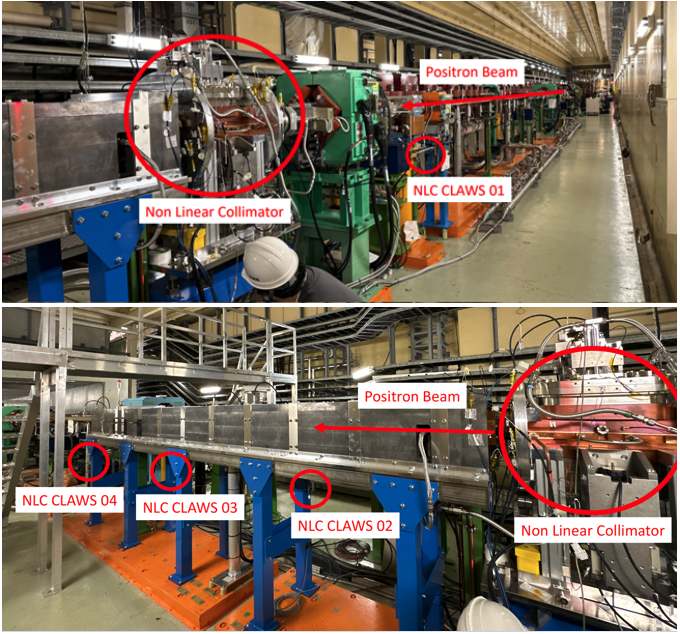}
\caption{D05 vertical collimator and CLAWS sensors. The vertical collimator is highlighted with the large red circle. It is called a Nonlinear Collimator. The position of the CLAWS sensors is highlighted with the small circles.}
\label{fig:CLAWSatD05V1}
\end{figure}

After careful consideration of factors such as reducing cable lengths and the strategy to utilize the newly constructed NLC collimator at D05 more actively for background reduction while minimizing collimator damage risks after LS1,
this collimator was selected as the first location for the additional CLAWS installation.
Consequently, four CLAWS sensors were installed around the D05V1 collimator as shown in Figure~\ref{fig:CLAWSatD05V1}. One sensor was positioned upstream of the collimator, while the remaining three were placed downstream.

\begin{figure}[htb]
\centering
\includegraphics[width=\linewidth]{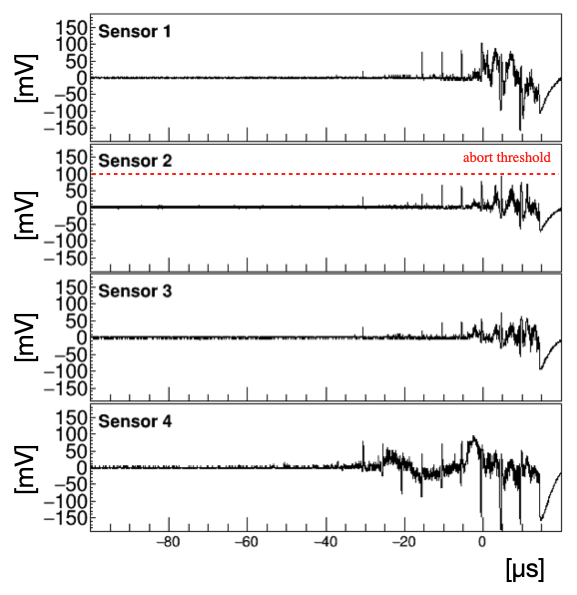}
\caption{Signals from the CLAWS sensors installed around the D05V1 collimator during a beam abort event. Sensor \#1 positioned upstream of the collimator and sensors \#2, \#3, and \#4 located downstream. The magnitude of the signals reflects the beam loss detected at each sensor, with Sensor \#4 (the most downstream sensor) generally recording the largest losses.}
\label{fig:d05claws_signal}
\end{figure}

During the operations in early 2024, the commissioning of the new CLAWS sensors was conducted. Figure~\ref{fig:d05claws_signal} illustrates a typical signal during a beam  abort. 

Sensors \#1 (upstream) and \#4 (most downstream) generally observed larger beam losses. In the case of SBL events, the beam loss around this collimator was significant, causing saturation in all CLAWS sensors. Sensor \#2 was selected as the abort sensor to ensure a broad threshold range for the abort request condition. Initially, the abort request threshold was set to 35~mV, which was later increased to 100~mV by the end of the 2024 spring run period. 

In addition to the threshold, the signal pulse width (duration) was incorporated into the abort request condition to effectively distinguish SBLs from other beam losses or noise. Commissioning results indicated that significant beam losses during SBL events last more than 1 $\mu$s. The abort request signal is triggered when the beam loss waveform exceeds a duration of 180~ns. An injection veto was also incorporated into the abort request logic of the CLAWS at D05. The launch of the abort request signal is vetoed for 1~ms after injection. This condition was optimized during the 2024 spring operation.

An integration study of CLAWS into the abort system was conducted using the above abort request conditions by providing pseudo-abort requests with the new CLAWS abort sensor and comparing its response time with that of the operational abort trigger system. Most of the time, the pseudo-abort request was faster than the operational abort trigger, depending on which sensor first launched the abort request. The test results clearly demonstrated the effectiveness of the new CLAWS sensor. During this commissioning period, only once did the pseudo-abort request launch alone, which could be considered an unnecessary abort request. However, given its very low frequency of once a week, it was deemed acceptable for beam operation and commissioning. After confirming the feasibility of the new CLAWS sensor, it was integrated as the abort sensor in the SuperKEKB LER by the end of March 2024.

Following the installation of the D05V1 CLAWS, a total of 168 SBL events were recorded in the LER during the 2024 spring operation~\footnote{The 168 SBL events include those that occurred during the physics run, when the Belle~II DAQ was operational, as well as during machine studies.}. In 148 of these instances, the D05V1 CLAWS generated the fastest abort request signal. Preliminary analyses suggest that the CLAWS abort request was issued 5-10~$\mu$s earlier than the second-fastest abort source. This improvement is anticipated to significantly mitigate beam loss and reduce potential damage to the IR components. Given the successful performance of the D05V1 CLAWS, the decision has been made to install additional CLAWS units around the D06V1 collimator during the 2024 summer shutdown.

\subsection{Expansion of Abort Modules}
To further speed up the abort response by optimizing the abort request signal path, the second master module of the abort trigger system was installed at the D07 power station during LS1. A direct optical line now connects D05 to D07, where the abort kicker power supply is located. With the master module installed at D07, it is possible to issue abort requests through this signal line without routing through the CCB. This modification changes the path from (2) to (3) in Figure~\ref{fig:fast_abort_scenarios}. In this configuration, the response time of the abort system is expected to be shortened by 1.4~$\mu$s for D05 and 1.9~$\mu$s for D06, respectively. In practice, the operation timing of the abort kicker magnet is synchronized with the abort gap and quantized at 5~$\mu$s intervals, so there is a 28\% (D05) and 38\% (D06) chance that the beam abort will be executed 5~$\mu$s faster.

\begin{figure}[htb]
\begin{center}
\includegraphics[width=78mm]{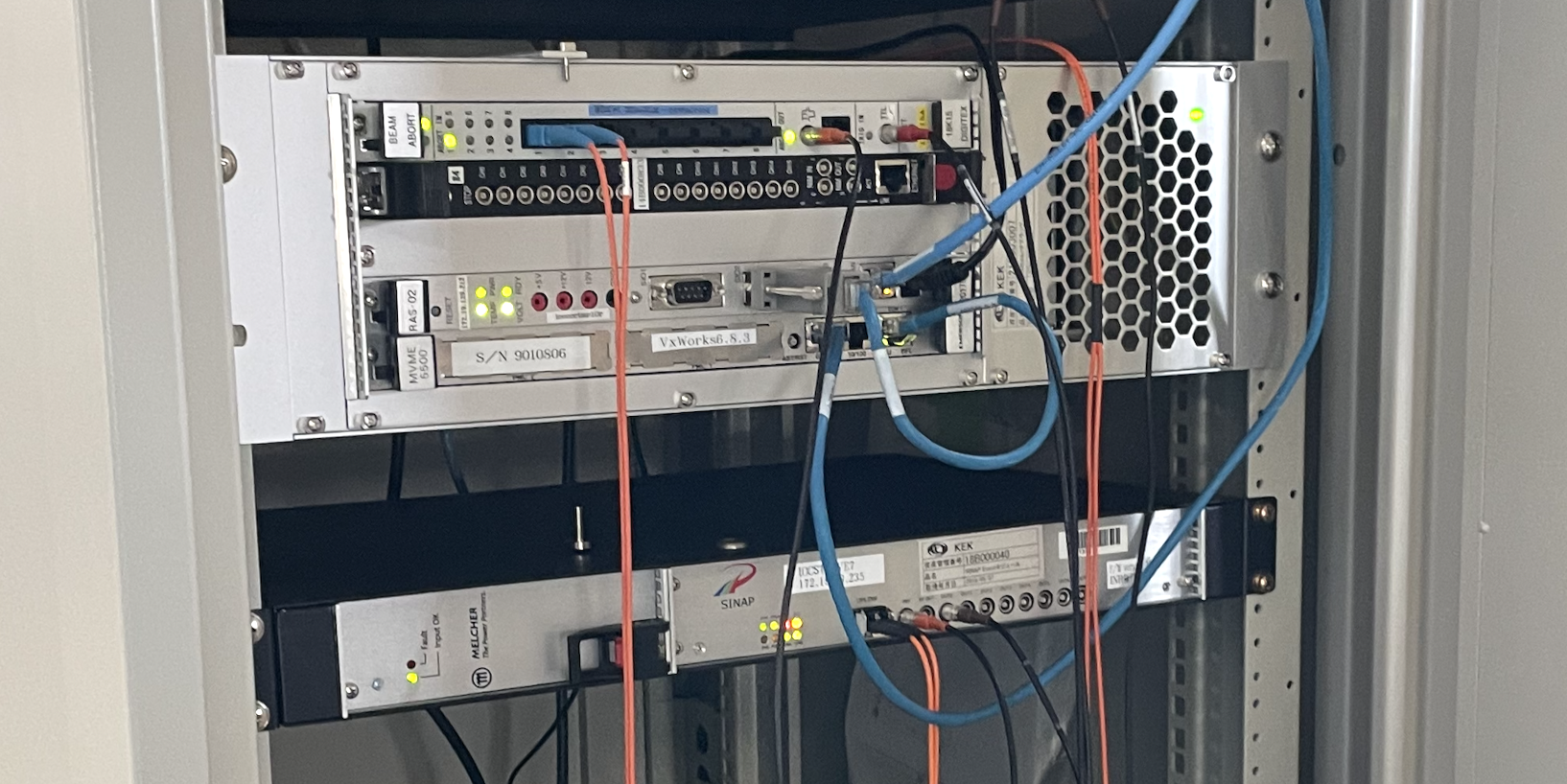}
\end{center}
\vspace{-5mm}
\caption{Second abort master module and abort kicker magnet trigger circuit. The abort master module is inserted in the top slot of the VME subrack, and the Event Timing System module is on the bottom.}
\label{fig:AbortTest}
\end{figure}

Figure~\ref{fig:AbortTest} shows the second abort trigger system. This system is equipped with the same functions as the master module and abort kicker magnet trigger circuit currently installed in the CCB. The master module is a device that aggregates abort request signals from the slave modules at each power station and outputs trigger signals to the abort kicker magnet trigger circuit. The master module uses the VME-type platform circuit board. The abort request signals input to the second master module are issued by the slave module (2-channel abort optical output circuits~\cite{AbortTrigger}) installed at the D05 power station, and it is connected to the newly installed CLAWS sensors.

The abort kicker magnet trigger signal is derived from a doubled signal of the accelerator’s beam-cycle signal, which corresponds to one beam-cycle period, and is delayed in synchronization with the abort gap. The Event Timing System module, newly installed at the D07 power station, functions as the output circuit for the turn signal, with one of its output terminals designated as the abort kicker magnet trigger. The abort request signal from the test master module is fed into the INHIBIT input terminal of the Event Timing System module, enabling precise control of the kicker trigger output.

This setup allows us to verify the speedup of the abort response by combining ``early detection of SBL in D05” and ``shortening the abort request signal path.” The verification test of this system commenced with the 2024 operation of SuperKEKB. Measurements were conducted over three days using the CLAWS installed at D05V1. For 14 abort events where the CLAWS at D05V1 issued the abort signal, it was confirmed in two cases that the beam abort was executed 5~$\mu$s faster when bypassing the CCB compared to when it was routed through it. This result validates the expected improvement in response time. Additionally, the abort system is expected to serve as a test environment for the laser abort system development described in the next section.

As a further improvement, plans are being developed to bypass the D05 power station and directly connect the CLAWS signal to the abort master module at D07. This adjustment will reduce the signal transmission path to 700~m and the corresponding signal transfer time to 3.5~$\mu$s. For comparison, the signal transfer time from the D05 power station to the D07 power station via the CCB is 6.1~$\mu$s~\footnote{The CLAWS installed at D05V1 is approximately 100~m away from the D05 power station. The signal transfer time within this 100~m distance is ignored in this estimation.}. This modification is expected to reduce the response time by at least 2.6~$\mu$s.

The situation is expected to improve further with the planned installation of CLAWS near the D06V1 collimator in summer 2024. The signal transfer time to the D07 power station will be reduced to 1.5~$\mu$s. Additionally, D06V1 is located approximately 400~m upstream of D05V1, which enhances the early detection of SBL events. While signal attenuation over this distance was initially a concern, tests have shown that the signal retains 10\% of its original amplitude after transferring 700~m through the CLAWS cable. This level of attenuation is considered acceptable, given the substantial beam loss signal associated with SBL events.

\subsection{Laser Abort System}
Studies have been conducted to explore the use of lasers as a new transfer technology for abort request signals. In the current system, the abort request signal is transmitted using light with a wavelength of 820~nm. The master-slave modules of the abort trigger system operate on a negative logic specification: they constantly transmit light during normal operation (no abnormalities), and the cessation of this light transmission is interpreted as an abort request signal. This simple yet robust technology has inspired further research into laser-based transmission methods.

\begin{figure}[htb]
\begin{center}
\includegraphics[width=78mm]{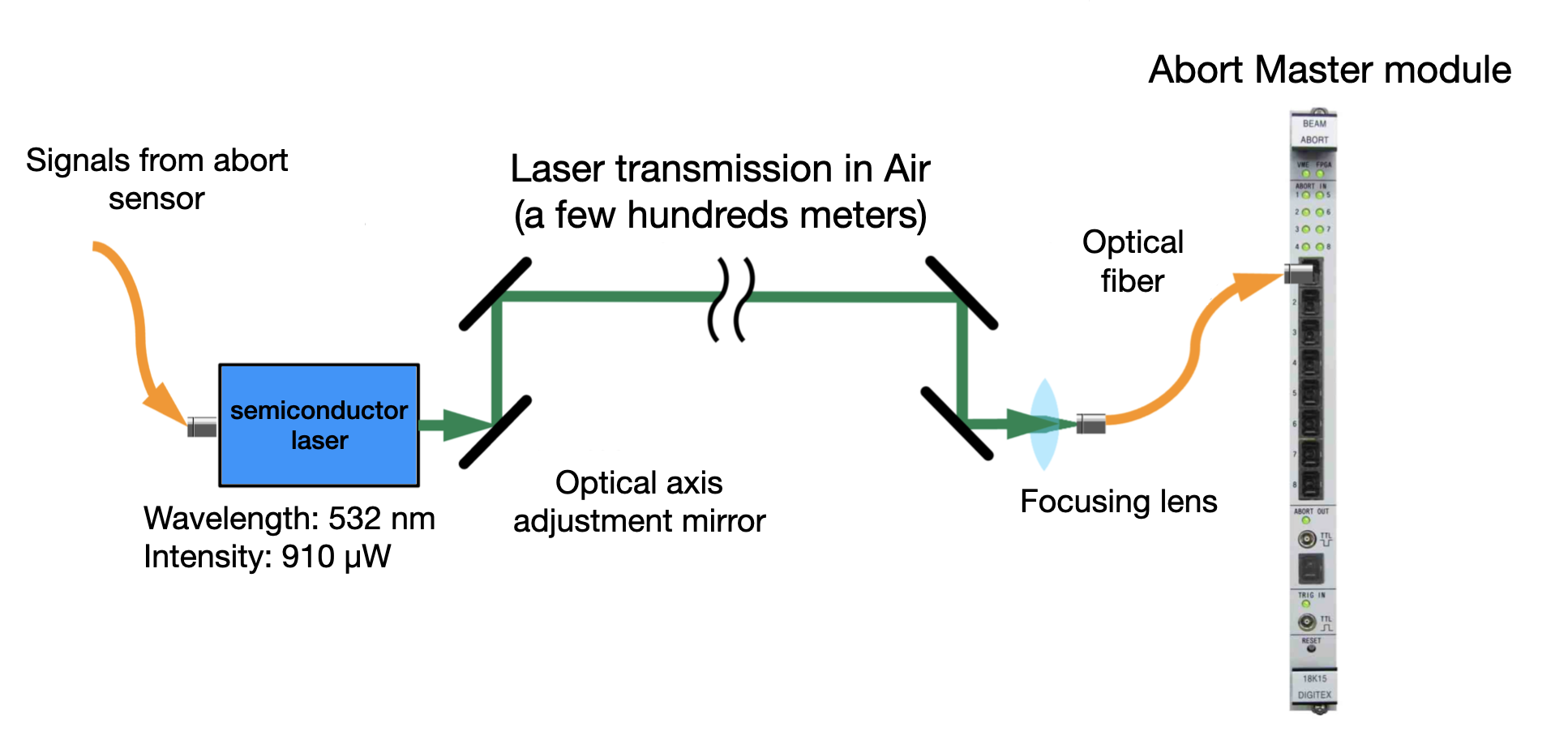}
\end{center}
\vspace{-5mm}
\caption{Setup for the laser-based abort signal transmission system. The system includes a semiconductor laser with a wavelength of 532~nm and an intensity of 910~$\mu$W. The laser signal is controlled by an abort sensor and transmitted through the air over a few hundred meters. Optical axis adjustment mirrors ensure the beam remains aligned, and the laser light is finally focused into an optical fiber using a focusing lens before being input into the abort master module.}
\label{fig:LaserAbort}
\end{figure}

Using a highly directional laser instead of optical fiber to transmit the light signal allows for faster transmission due to the lower refractive index of air. Light can travel 1.5 times faster in air (refractive index of about 1) than in quartz, which is commonly used in optical fibers (refractive index of about 1.5). The transmission time difference between using optical fiber and laser to send light from the D05 and D06 areas of the accelerator tunnel to the D07 area is about 0.9~$\mu$s and 0.7~$\mu$s, respectively. This research also looks ahead to next-generation collider experiments, which may feature even longer beamlines than SuperKEKB, making reduced transmission time even more beneficial. Figure~\ref{fig:LaserAbort} shows a conceptual diagram of the laser abort request signal system. The signal from abort sensors, such as CLAWS, is used to control the power supply of a diode laser, enabling the laser to be toggled ON and OFF. The transmitted laser light is then focused into an optical fiber using a focusing lens positioned near the abort master module, where it is subsequently injected into the module.


A class~II green laser with a wavelength of 532~nm was selected as the signal source for several key reasons. Firstly, the abort trigger module is highly responsive to light in the green wavelength range. Secondly, successful operation of the abort master module, with a laser input power of 30~$\mu$W, has been demonstrated through controlled laser activation. Additionally, the use of visible laser light simplifies the maintenance of the optical path. Finally, there are no regulatory restrictions on the operation of class~II lasers, further supporting their suitability for this application.

\begin{figure}[htb]
\begin{center}
\includegraphics[width=78mm]{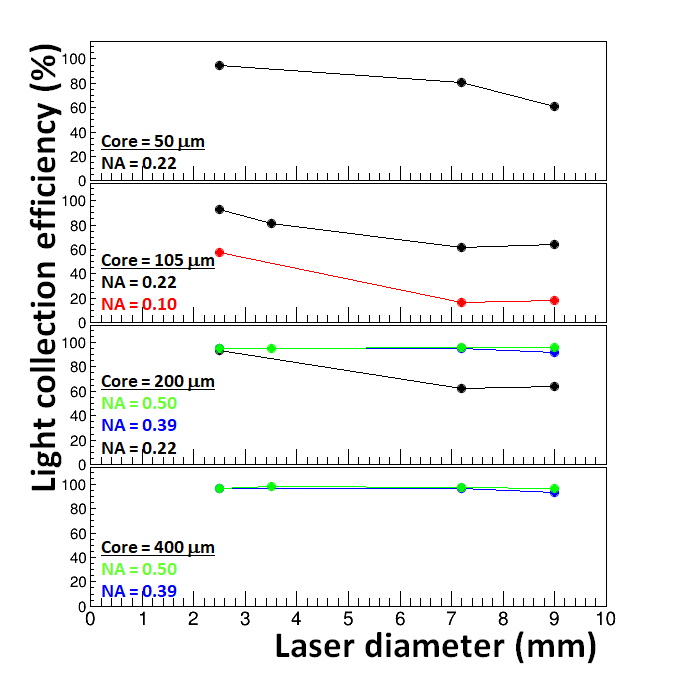}
\end{center}
\vspace{-5mm}
\caption{Light collection efficiency for various laser diameters across different types of optical fibers, with cores of 50~$\mu$m, 105~$\mu$m, 200~$\mu$m, and 400~$\mu$m, and numerical apertures (NA) ranging from 0.10 to 0.50. The results illustrate the impact of increasing laser diameter on the efficiency of light collection, which generally decreases as the laser diameter increases, particularly in fibers with smaller cores and lower NAs.}
\label{fig:coll_eff}
\end{figure}

The technology to efficiently focus the transported laser into the fiber has been well established~\cite{kitamura2023}, achieving a light collection efficiency exceeding 90\%, while only 3\% efficiency is required to collect the 30~$\mu$W light necessary for operation. The collection efficiency has been measured relative to the laser diameters for various types of optical fibers, each with different core diameters and numerical apertures (See Figure \ref{fig:coll_eff}). By selecting the appropriate optical fiber, such as one with a large core diameter or a high numerical aperture, the input requirements for the abort master module can be reliably met.

Another crucial aspect of the laser abort signal system is the optical technology required to maintain the laser beam profile during long-distance transmission. Typically, the beam size broadens over long distances due to imperfect beam collimation and light scattering by air molecules. As shown in Figure~\ref{fig:coll_eff}, there is a noticeable decrease in efficiency as the laser diameter increases.

Two experimental setups were developed, each involving the transmission of a 910~$\mu$W laser over a 400m distance to evaluate the system’s feasibility~\cite{kitamura2023}. The first setup used a telescope configuration with two convex lenses to create a parallel beam, where the focal lengths of the lenses were a few tens of centimeters. With this setup, a light intensity of 581~$\mu$W was successfully collected in the optical fiber. The second setup employed spherical mirrors with focal lengths of a few tens of meters. This configuration resulted in a slight improvement in light intensity, achieving 642~$\mu$W. In both cases, the resulting light intensity was nearly 20 times the required threshold of 30~$\mu$W. This was achieved through careful optimization of the optical systems used in the setups.

\begin{figure}[htb]
\begin{center}
\includegraphics[width=78mm]{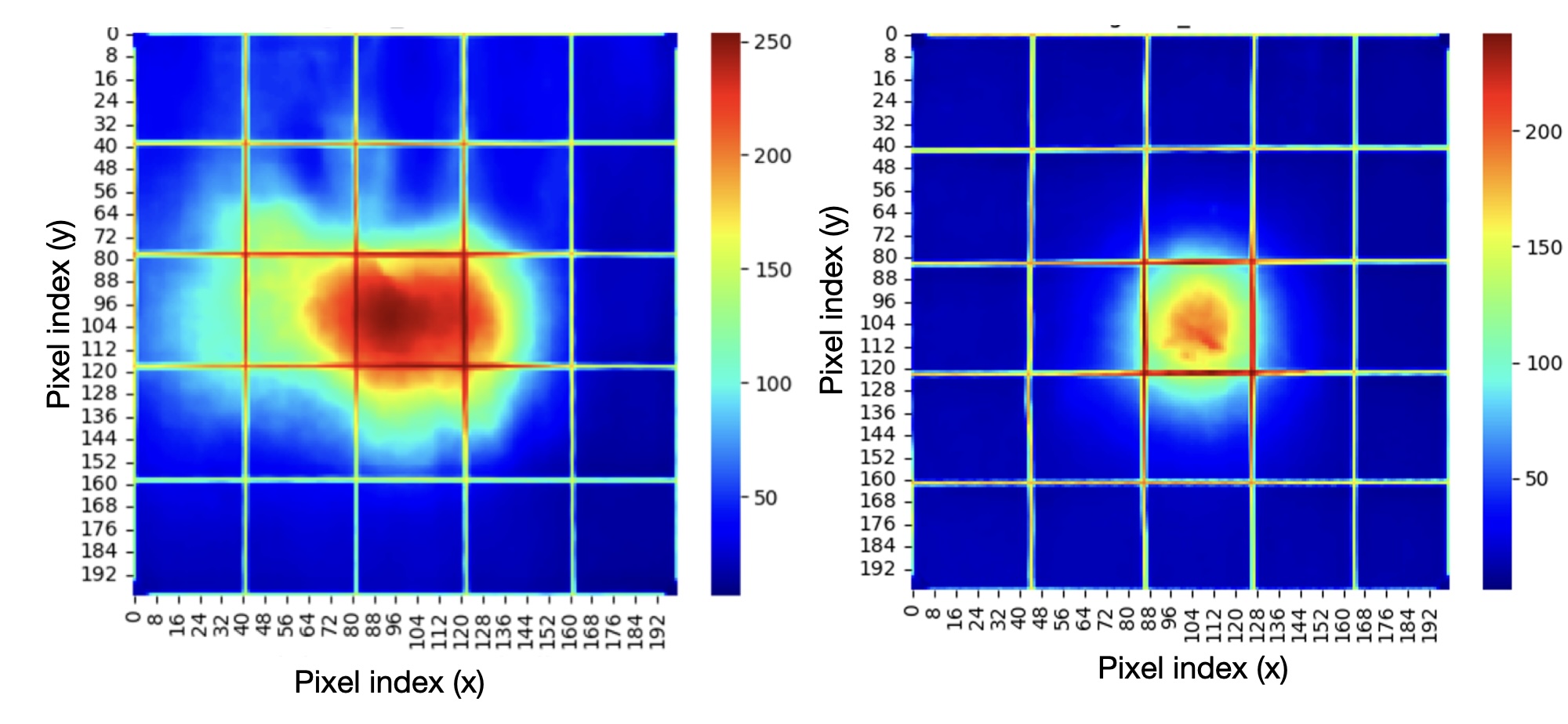}
\end{center}
\vspace{-5mm}
\caption{Comparison of beam sizes: Convex lens focusing (left) and spherical mirror focusing (right) over a 400~m distance. The Z axis represents the total number of events accumulated over time.}
\label{fig:BeamSizeComparison}
\end{figure}

By using spherical mirrors, not only was the light intensity improved, but the beam size was also reduced, resulting in a more compact and symmetric profile, as shown in Figure~\ref{fig:BeamSizeComparison}. The relatively degraded profile observed with the convex lens optics may be attributed to the challenges in aligning the two lenses. The impact of misalignment becomes more pronounced when transmitting over long distances. This result highlights the effectiveness of spherical mirrors in maintaining beam quality over extended distances. Additionally, when conducting experiments in the ring tunnel, reflection mirrors are necessary. Spherical mirrors, which provide both focusing and reflection capabilities, reduce the number of components required compared to the use of convex lenses.

Following the successful demonstration in the above two experimental setups,
further tests were conducted within the SuperKEKB tunnel. The tests involved a 244~m laser transmission from D06V1 to the boundary of D06/D07 using spherical mirrors to maintain beam quality. These tests aimed to confirm the system’s performance in a realistic operational environment. The laser was transmitted across multiple optical setups mounted on cable racks within the tunnel, achieving stable transmission and efficient light collection. The tunnel experiments demonstrated that the laser beam profile remained circular, similar to the previous experiments in the LINAC area.

\begin{figure}[htb]
\begin{center}
\includegraphics[width=78mm]{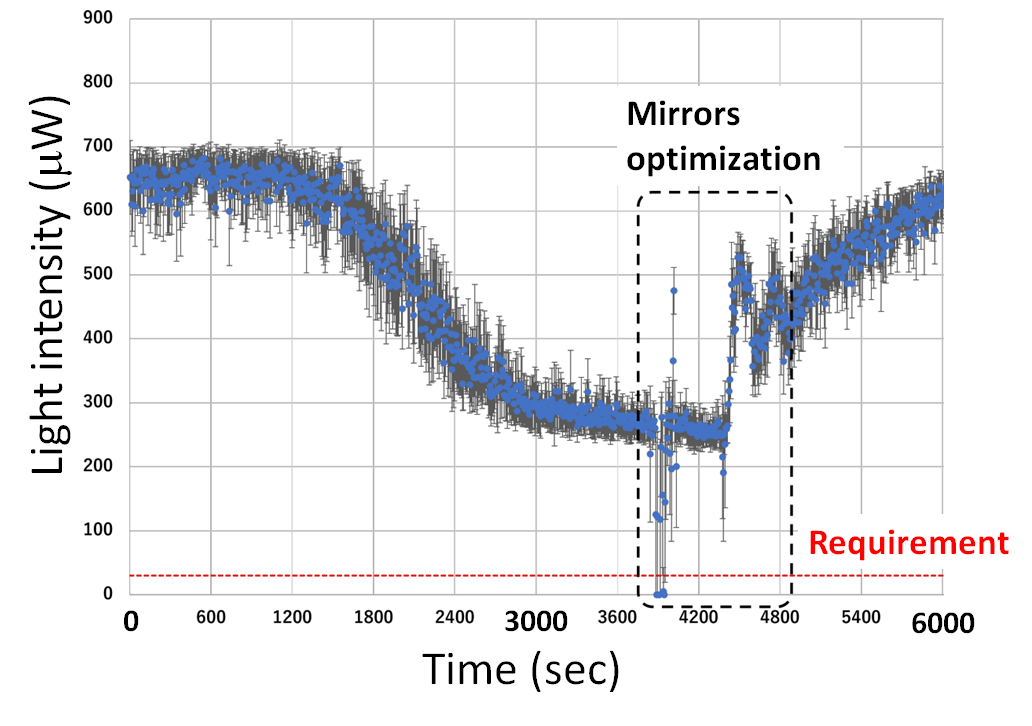}
\end{center}
\vspace{-5mm}
\caption{Long-term measurement of laser intensity within the SuperKEKB tunnel. The fluctuations in intensity are shown over time, with a brief dip observed during mirror optimization. The system recovers and consistently exceeds the required threshold of 30 $\mu$W for the abort master module operation.}
\label{fig:skb_result}
\end{figure}

Figure~\ref{fig:skb_result} shows the time-dependent laser intensity measured within the SuperKEKB tunnel. While fluctuations in intensity were observed over extended periods, the system consistently met the requirement of 30 $\mu$W, which is the threshold for ensuring proper operation of the abort master module. The observed degradation in intensity may be attributed to the drift in the laser’s optical path. This drift is likely caused by temperature variations within the tunnel, leading to the deformation of the frame that holds the laser optics in place. Such deformation could alter the mirror angles and subsequently the laser’s trajectory. Supporting this hypothesis, the laser intensity was restored by adjusting the mirrors’ angles. These findings demonstrate the potential viability of utilizing a laser-based abort signal system in the SuperKEKB tunnel, particularly if a remote mirror angle control system is implemented.


This innovative approach is expected to significantly enhance the speed of the abort response system, thus mitigating the damage caused by SBL and contributing to the overall efficiency and safety of the SuperKEKB accelerator.


\section{Future Directions}

To further advance the understanding and mitigation of SBL at SuperKEKB, we are focusing on two critical areas of beam diagnostics: Bunch Oscillation Recorder (BOR) and X-ray Beam Size Monitor (XRM). These instruments will provide valuable data on beam oscillations and beam size variations, respectively, allowing for a more comprehensive analysis of SBL events.

\subsection{Bunch Oscillation Recorder (BOR)}
In our new beam diagnostic system, beam losses detected by the loss monitors installed around the collimators have been the primary focus to investigate the causes of SBL events. However, upon examining beam oscillations recorded by BOR, it has been found that in many cases, oscillations begin several beam-cycles before the actual losses occur. 

The BOR is designed to record the beam position bunch-by-bunch for multiple turns before a beam abort. Before LS1, a single BOR~\cite{BORBCM} was installed on each ring, primarily for the Bunch Feedback System monitor, but this setup was insufficient to pinpoint the initial cause and location of SBL events. 

Moving forward, it will be crucial to acquire such oscillation data at various points around the ring. By further analyzing the oscillation data and identifying the locations where these oscillations start, we can potentially gain more insights into the causes of SBL. 

\begin{figure}[htb]
\centering
\includegraphics[width=78mm]{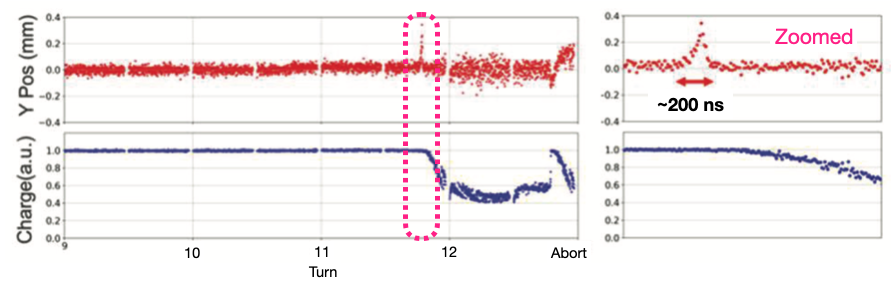}
\caption{SBL event recorded with RFSoC BOR: vertical position (upper) and charge (lower) for the last four turns before the abort.}
\label{fig:BOR_signal}
\end{figure}

As the first step towards this goal, we have developed a portable BOR~\cite{RFSoCBOR} using Radio Frequency System on Chip (RFSoC) evaluation board ZCU111 from AMD/Xilinx in addition to the existing BOR. The RFSoC-based BOR will allow for multiple installations around the ring to cover a larger range of phase advances and detect bunch oscillations more accurately. Figure~\ref{fig:BOR_signal} shows an SBL event recorded with the new RFSoC BOR.

\subsection{X-ray Beam Size Monitor (XRM)}
The X-ray Beam Size Monitor (XRM)~\cite{XRM} at SuperKEKB is designed to measure the vertical beam size using synchrotron radiation from bending magnets. The XRMs are installed one to each main ring. The current setup employs optical elements such as pinholes and coded apertures, along with a scintillator screen and a CMOS camera, for high-resolution multi-bunch measurements. 

In addition, a high-speed Si-based XRM is under development to enable single-shot (single bunch, single turn) measurements, providing bunch-by-bunch vertical beam size data. This advanced system will use a deep Si detector and high-speed readout electronics with coded aperture imaging to enhance photon throughput. The new XRM will significantly improve the identification of SBL causes by allowing detailed bunch-by-bunch measurements at multiple ring locations. This capability is expected to play a crucial role in analyzing beam instabilities and enhancing low-emittance tuning for better overall accelerator performance.

\section{Conclusion}
Significant strides have been made in developing and implementing advanced beam diagnostic and abort systems to address the challenges posed by SBL events at SuperKEKB. The high-speed beam diagnostic system with improved time resolution, which includes high-speed loss monitors and the WR time synchronization system, has greatly improved our ability to locate the onset of beam losses within the SuperKEKB ring. The installation of additional CLAWS detectors and the expansion of abort modules have demonstrated potential for lessening the impact of beam losses on Belle~II by reducing the response time of the abort system.

The possible mechanisms behind SBL events have been explored, such as beam-dust interactions and the fireball hypothesis, and acoustic sensors have been developed. Although acoustic sensors have not yet detected discharge phenomena near collimators, they remain a promising tool for future studies. Our efforts in understanding these mechanisms are ongoing, and further research is necessary to fully elucidate the causes of SBL. The development of the BOR and the XRM is expected to provide critical data, enhancing our ability to diagnose and mitigate SBL events more effectively.

The advancements made in beam diagnostics and abort systems not only aim to minimize the risks associated with SBL events but also support the overarching goal of achieving higher luminosities at SuperKEKB. These improvements will enable the Belle~II experiment to collect high-statistics data essential for exploring new physics beyond the Standard Model. The success of these initiatives reflects the collaborative efforts of the SuperKEKB and Belle II teams, and we look forward to continued progress in enhancing the performance and stability of the accelerator.

\section*{Acknowledgements}
We express our gratitude to the SuperKEKB accelerator team and all collaborators of the SuperKEKB and Belle~II groups for their tremendous cooperation in achieving stable machine operation and aiding in the development of the beam diagnostic system and advanced beam abort system. This progress has been made possible through the participation of many graduate students from Japan and abroad in the Belle II group, as well as the understanding and support of the supervising professors regarding Machine Detector Interface (MDI) projects. 

We also extend our sincere thanks to C.~Niebuhr (DESY) and H.~Fukuma (KEK) for their valuable review and discussion regarding this paper.

This work was supported by the Ministry of Education, Culture, Sports, Science, and Technology (MEXT) of Japan, KAKENHI (Grant Numbers 22H03867, 22K21347, and 23H05433), and the US-Japan Ozaki Exchange Program~\cite{OzakiProgram}.


\section*{Declaration of generative AI and AI-assisted technologies in the writing process}
During the preparation of this work the author(s) used ChatGPT in order to assist with language polishing. After using this tool/service, the author(s) reviewed and edited the content as needed and take(s) full responsibility for the content of the publication.

\bibliographystyle{elsarticle-num}
\bibliography{references}

\end{document}